\documentclass[twocolumn,usenatbib]{mnras}

\usepackage{newtxtext,newtxmath}
\usepackage{esvect}
\usepackage[font=small,skip=0pt]{caption}

\usepackage{graphicx}
\usepackage{appendix}
\usepackage{float}
\title[Radial Gas Flows on FIRE]{Gas infall and radial transport in cosmological simulations of Milky Way-mass disks}

\author[C. Trapp et al.]{
Cameron W. Trapp$^{1}$,\thanks{E-mail: ctrapp@ucsd.edu}
Du\v{s}an Kere\v{s}$^{1}$,
Tsang Keung Chan$^{1,2}$,
Ivanna Escala$^{3,4}$\thanks{Carnegie-Princeton Fellow},
\newauthor
Cameron Hummels$^{5}$,
Philip F. Hopkins$^{5}$,
Claude-Andr\'e Faucher-Gigu\`ere$^{6}$,
Norman Murray$^{7,8}$,
\newauthor
Eliot Quataert$^{3}$,
and Andrew Wetzel$^{9}$
\\
$^{1}$Center for Astrophysics and Space Sciences (CASS), University of California San Diego, 9500 Gilman Dr, La Jolla 92093, USA\\
$^{2}$Institute for Computational Cosmology, Durham University, South Road, Durham DH1 3LE, UK\\
$^{3}$Department of Astrophysical Sciences, Princeton University, Princeton, NJ 08544, USA\\
$^{4}$The Observatories of the Carnegie Institution for Science, 813 Santa Barbara St, Pasadena, CA 91101, USA\\
$^{5}$TAPIR, Mailcode 350-17, California Institute of Technology, Pasadena, CA 91125, USA\\
$^{6}$Department of Physics and Astronomy and CIERA, Northwestern University, 1800 Sherman Ave, Evanston, IL 60201\\
$^{7}$Canadian Institute for Theoretical Astrophysics, 60 St. George Street, University of Toronto, ON M5S 3H8, Canada\\
$^{8}$Canada Research Chair in Astrophysics\\
$^{9}$Department of Physics and Astronomy, University of California, Davis, CA 95616\\
}

\date{Accepted XXX. Received YYY; in original form ZZZ}

\pubyear{2020}

\newcommand{\tgal}{$t_{\rm orbit}$ }
\newcommand{\rgal}{$R_{\rm DLA}$ }
\newcommand{\rgalStop}{$R_{\rm DLA}$}
\newcommand{\hgal}{$h_{\rm total}$}
\newcommand{\msun}{ \rm M_{\odot}}
\newcommand{\MFunit}{M$_{\odot}$ yr$^{-1}$}

\begin{document}
\label{firstpage}
\pagerange{\pageref{firstpage}--\pageref{lastpage}}

\maketitle

\begin{abstract}

Observations indicate that a continuous supply of gas is needed to maintain observed star formation rates in large, disky galaxies. To fuel star formation, gas must reach the inner regions of such galaxies. Despite its crucial importance for galaxy evolution, how and where gas joins galaxies is poorly constrained observationally and rarely explored in fully cosmological simulations. To investigate gas accretion in the vicinity of galaxies at low redshift, we analyze the FIRE-2 cosmological zoom-in simulations for 4 Milky Way mass galaxies ($M_{\rm halo}\sim10^{12}M_{\odot}$), focusing on simulations with cosmic ray physics. We find that at $z\sim0$, gas approaches the disk with angular momentum similar to the gaseous disk edge and low radial velocities, piling-up near the edge and settling into full rotational support. Accreting gas moves predominately parallel to the disk and joins largely in the outskirts. Immediately prior to joining the disk, trajectories briefly become more vertical on average. Within the disk, gas motion is complex, being dominated by spiral arm induced oscillations and feedback. However, time and azimuthal averages show slow net radial infall with transport speeds of 1-3~km~s$^{-1}$ and net mass fluxes through the disk of $\sim$M$_{\odot}$ yr$^{-1}$, comparable to the galaxies' star formation rates and decreasing towards galactic center as gas is sunk into star formation. These rates are slightly higher in simulations without cosmic rays (1-7~km~s$^{-1}$, $\sim$4-5~M$_{\odot}$ yr$^{-1}$). We find overall consistency of our results with observational constraints and discuss prospects of future observations of gas flows in and around galaxies.

\end{abstract}

\begin{keywords}
galaxies: evolution -- stars: formation -- galaxies: kinematics and dynamics -- galaxies: spiral
\end{keywords}

\graphicspath{{./}{Figures/}}

\section{Introduction} \label{sec:intro}


Massive disk galaxies, including our own Milky Way, show relatively stable star formation rates (SFRs) over the past few Gyr \citep{binney00} and actively form stars over cosmic time. Such star formation cannot be solely supplied by existing gas reservoirs in the inter-stellar medium (ISM), as measured gas depletion times are too short for sustained star formation. Typical depletion times of molecular gas  ($t_{\rm dep}=\rm M_{\rm H_{2}}/\rm{SFR}$) are $\sim$1-2 Gyr at present time and even shorter at higher redshift \citep{tacconi18,saintonge17}. In large, disky galaxies such as the Milky Way (MW), more extended neutral hydrogen reservoirs can condense to H$_{2}$ and continue fueling star formation, but these would still deplete within $\sim$2 Gyrs \citep{kennicutt98}. Given that these depletion times are an order of magnitude shorter than the Hubble time ($\sim$14 Gyrs), these reservoirs must be resupplied over time. Gas recycling via stellar mass loss \citep{leitner11} can provide a partial source, but a continuous supply of gas is still necessary to maintain observed SFRs.  

More local observations, such as the G-Dwarf problem (i.e. the relative scarcity of low metallicity stars in the solar vicinity does not match predictions from simple galactic chemical evolution models) \citep{bergh62,schmidt63,sommer-larsen91,worthey96,haywood19}, additionally motivate the need for continuous accretion of low metallicity gas from the circumgalactic medium (CGM) in present day disks.

   
One of the most direct ways to account for this accretion is through observation of Intermediate- and High-Velocity Clouds (IVCs and HVCs). 
IVCs and HVCs are gaseous clouds with strong kinematic deviations from galactic rotation, so this classification selects gas that is not yet part of the rotationally supported disk. These deviations are typically within 40-70 km/s for IVCs, and above 70-90 km/s for HVCs \citep{Rohser18}. Observations of HVCs around the MW have shown total gas accretion rates of around 0.4 $\rm M_{\odot}$/yr, which is not enough to fully support observed SFRs of $\sim$2-3$\rm M_{\odot}$/yr \citep{putman12}. IVC accretion is likely related to galactic fountain recycling \citep{putman12} and may provide a significant fraction of total accretion \citep{Rohser18}, however, further observational studies are required.

The nature of gas accretion over cosmic time from the inter-galactic medium to galactic regions has been extensively studied in hydrodynamic galaxy formation simulations. Broadly speaking, cosmological simulations show that star formation is largely supply driven  and typical accretion rates are on the order of galactic star formation rates \citep[e.g.][]{keres05}. Additionally, simulations found that gas can accrete in "cold-mode", along filamentary streams, where gas does not shock to the virial temperature in the outer halo \citep{keres05, dekel06, ocvirk08, brooks09, keres09a,faucher11, vandevoort11a,stern20b}. Such accretion dominates at early times, while at late times and/or in more massive halos cooling from the hot halo gas ("hot-mode") can be more important. The present-day Milky Way is expected to be in the "hot mode" regime, but close to the mass where the transition between these two cases occurs \citep{keres05,dekel06, ocvirk08, vandevoort11b, nelson13,stern20a}, and it is therefore especially interesting. Furthermore, when stellar feedback and resulting galactic outflows are included, a significant fraction of  late time accretion originates from previous episodes of gas ejection from galaxies \citep{oppenheimer10, wetzel15, muratov15,alcazar17, hafen17}. Further investigation of accretion in the context of cold mode/hot mode in the FIRE simulations is the subject of upcoming work \citep{hafen_inPrep}.

Simulations also suggest that high-redshift gas accretion can co-rotate with the disks near the edge of the stellar disk \citep{keres05, Danovich15, stewart17}. However, high accretion rates and the resulting strong stellar feedback and outflows lead to relatively chaotic disks with high velocity dispersion and strong time variations in infalling gas \citep[e.g.][]{muratov15}.  This picture changes at later times in Milky Way-mass galaxies: infall is steady while outflows are weaker, facilitating formation of coherent disks  \citep[e.g.][]{kassin12, muratov15,stern20b}. This enables characterization of the gas infall with respect to the disk plane.

 Numerical simulations show that accreting cold gas is largely corotating at late times \citep{keres09b, stewart11b, ho19}.  Given the large specific angular momentum of overall halo gas \citep{elbadry18}, accreting gas can settle into rotational support in the disk outskirts regardless of its temperature history. This is consistent with the standard picture, where galactic disks grow inside-out \citep{fall80}
\footnote{If halo gas has significant non-thermal pressure support from cosmic-rays, infall geometry might be even more confined to galactic plane \citep{hopkins21_crOutflows}.}.
There is also indirect observational support for this scenario from larger scale gas flows, where cold/warm absorbers in halos of low redshift galaxies show co-rotation with the disk, potentially mapping such infalling gas \citep{bielby17,peroux17,diamond16,muzahid16}. Studies at higher redshifts utilizing background quasars to probe disk outskirts have also shown kinematic evidence for co-rotating structures out to 30-60 kpc \citep{barcons95,bouche13,zabl19}.

If this is the main way in which our own Galaxy accretes gas, it makes it hard to kinematically distinguish accreting gas from the gas already in the disk. This also suggests that a large fraction of accretion is not in HVCs, explaining the fact that  HVC accretion rates are much lower than what is needed to fuel observed SFR. Several studies have explored inflow from mostly co-rotating gas, though interpreting observed gas kinematics and inflow solely from the observed Doppler shift is difficult and can lead to large uncertainties \citep{wong04,martin12,rubin12, ho20}. Robust theoretical predictions for the nature of gas infall close to galaxies are clearly needed to test this scenario and guide future observations.



While our expectations are that gas will join the disk in the outskirts, most of the galactic star formation occurs in the inner regions. This means that an efficient transport mechanism is needed for gas to move from disk outskirts inward. Several theoretical studies have explored radial gas transport within the disk \citep[e.g.][]{krumholz10,dekel09,krumholz18,forbes19} and the radial migration of clumps through the disk at higher redshifts in cosmological simulations \citep[e.g.][]{ceverino10,mandelker14,hopkins12,oklopcic17}. However, radial transport velocity and mass flux of gas have not been explored in great detail in fully cosmological simulations of disk galaxies. On the other hand, observational studies searching for the radial gas flows within the disk are rare and inconclusive \citep{wong04,schmidt16}.

\begin{figure*}
    \center{\includegraphics[width=\textwidth]
	       {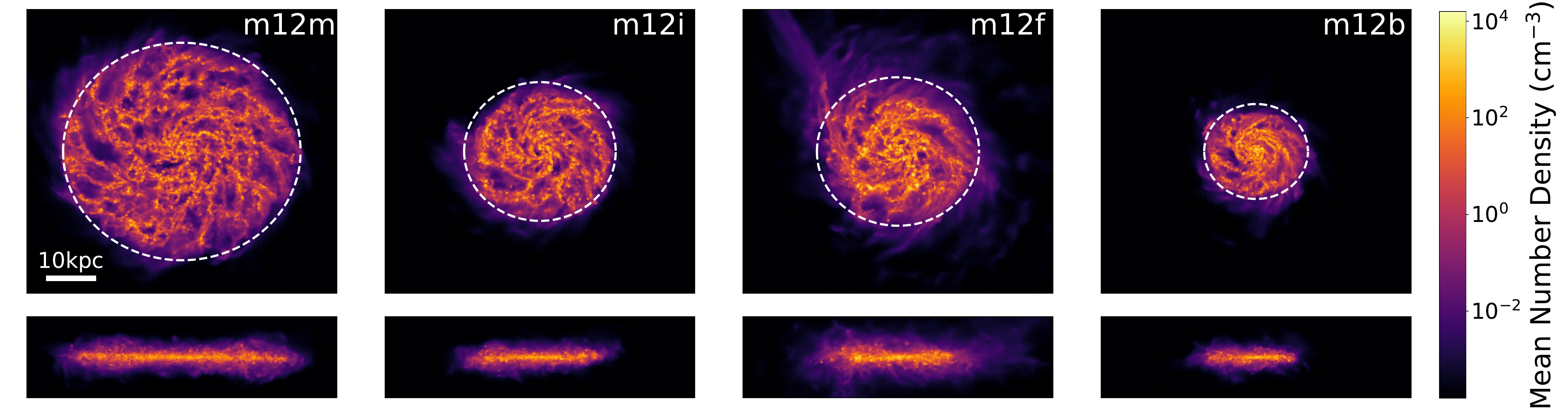}}
	  \caption{\label{fig:ba_dens} Mean gas density along line of sight for the 4 galaxies in our sample at redshift z=0. White dashed lines show the radius where the neutral hydrogen column density drops below 10$^{20.3} \rm{cm}^{-2}$ (\rgal, see Table~\ref{table:disks}). All four galaxies exhibit clear, thin gaseous disks. The disk in \textbf{m12f} underwent a recent merger event with an LMC like object at redshift~0.08, resulting in the streaming accretion feature in the top left quadrant and affecting numerous metrics in our analysis.
}
\end{figure*}
	
\begin{figure*}
	   \center{\includegraphics[width=\textwidth]
	       {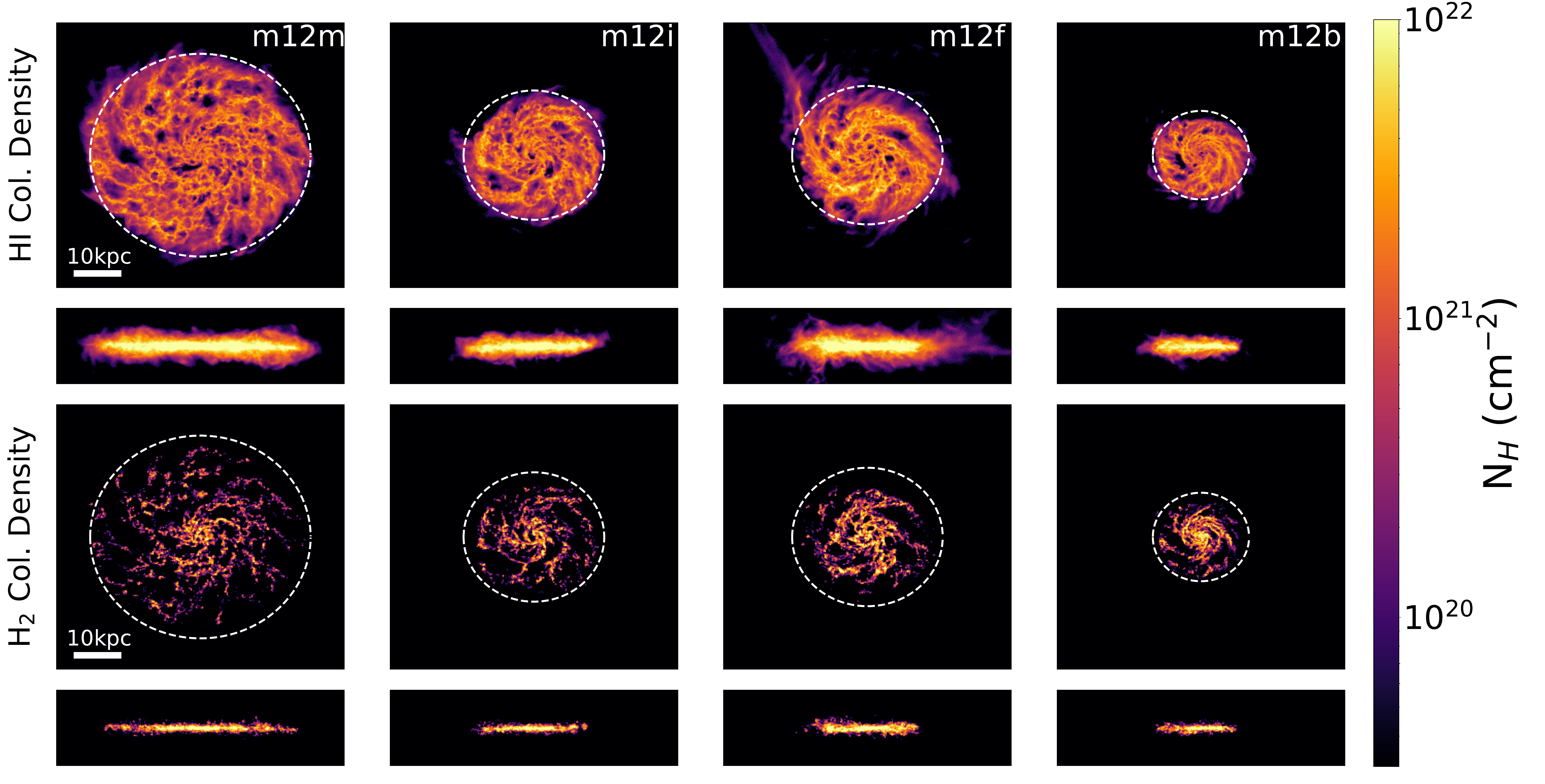}}
	  \caption{\label{fig:ba_coldens} Neutral (HI) and molecular (H$_{2}$) hydrogen column densities for the four galaxies in our sample. The white dashed lines show the radius of the disk (R$_{\rm DLA}$, Table~\ref{table:disks}). HI drops off sharply outside the disk. Molecular hydrogen is more concentrated towards the galactic center.
}
\end{figure*}

Our goal in this paper is to clarify the theoretical picture of galactic gas infall and radial gas transport in Milky Way-mass galaxies at late times and  provide predictions to guide future observations of these processes.
For this purpose, we utilize outputs from a subset of new FIRE-2 (Feedback In Realistic Environments) \footnote{http://fire.northwestern.edu} cosmological zoom-in simulations, which in addition to explicit stellar feedback models, incorporate magnetohydrodynamics and cosmic ray physics (CR+; \citealt{chan19,hopkins20}). A key parameter in these simulations--the effective diffusion coefficient of cosmic rays--was previously constrained based on observed galactic gamma ray emission \citep{lacki11}. We focus on $0 < z < 0.2$; this is the time when stable, thin, disks with clear orientation exist, enabling us to make robust predictions both externally and from the perspective within the disk. We characterize (i) the net accretion rates, star formation rates, and radial gas transport rates within the disk. We also characterize (ii) the directionality (i.e. parallel to the disk) of gas flow as it accretes onto the disk, (iii) the evolution of angular momentum of gas particles as they accrete, (iv) the radius at which particles join the disk, and (v) what happens to particles when they join the disk. Finally, we discuss the viability of observing these flows and compare these results with the hydrodynamical only (Hydro+) FIRE-2 runs and runs with magnetohydrodynamics but no CR physics (MHD+). CR+ runs were chosen for the main analysis due to their more realistic late time star formation compared to Hydro+ and MHD+ runs \citep{hopkins20}. A more detailed analysis of the phase evolution of accreting and transporting gas will be done in future work.

Section \ref{sec:simulations} describes the simulations used in detail. Section \ref{sec:analysis} describes the processing techniques used to analyze the data. Section \ref{sec:results} provides the key results from our simulation analysis. Section \ref{sec:discussion} discusses the significance of our findings and  introduces basic observational predictions. Appendix \ref{sec:appendix_comparisons} compares the CR+ runs presented in the main paper with the Hydro+ and MHD+ runs, Appendix~\ref{sec:appendix_II} contains supplementary figures to the main text, and Appendix ~\ref{sec:appendix_res_comp} compares simulations run at different mass resolution.

\section{Simulations}\label{sec:simulations}

\begin{table*}
\centering
\begin{tabular}{p{1.25cm} p{1.25cm} p{1.25cm} p{1.25cm} p{1.25cm} p{1.25cm} p{1.25cm} p{1.25cm} p{1.25cm} p{1.25cm} }
\hline
Simulation & $R_{\rm vir}$ & $R_{*,1/2}$ & \rgal &  $R_{\rm HI,19}$  & $h_{\rm total}$ & $h_{\rm{cold,inner}}$ & \tgal & $v_{\rm c}$ & $M_{*}$ \\
Name & [kpc] & [kpc] & [kpc] & [kpc] & [kpc] & [kpc] & [Gyr] & [km s$^{-1}$] & [$M_{\odot}$] \\
\hline
\hline
\textbf{m12m} & 232.0 & 7.84 & 26.8  &  30.5  & 0.87 & 0.21 & 0.874 & 190 & 3e10\\
\textbf{m12i} & 215.4 & 3.61 & 17.1 &  20.6 & 0.76 & 0.18 & 0.643 & 178 & 3e10\\
\textbf{m12f} & 237.1 & 3.72 & 18.3 &  28.8 & 1.01 & 0.33 & 0.675 & 193 & 4e10\\
\textbf{m12b} & 221.2 & 1.81 & 11.7 &  15.1 & 0.47 & 0.14 & 0.438 & 206 & 4e10\\
\hline
\hline

\end{tabular}

\caption{\label{table:disks} Parameters characterizing the size of the disk for the four galaxies in our sample at z=0. $R_{\rm vir}$ is the virial radius (calculated following \citet{bryan98}). $R_{*,1/2}$ is the radius at which half the stellar mass is contained. \rgal is the radius at which the total hydrogen column density drops below $10^{20.3} \rm cm^{-2}$ when viewed face on, signifying the transition to a column density below a Damped Lyman Alpha (DLA) system. Likewise, $R_{\rm HI,19}$ is the radius at which the HI column density drops below $10^{19} \rm cm^{-2}$. $h_{\rm total}$ is the scale height of the total gas and $h_{\rm cold,inner}$ is the scale height of the cold hydrogen ($T <$ 100 K) in the inner 5 kpc. Scale height was calculated as the height the average gas density drops by a factor of \textit{e} from the average value within $\pm$20 pc of the midplane. The parameter \tgal is the dynamical time of the galaxy, defined as the average orbital period of disk gas within a 1 kpc bin centered at \rgalStop. The rotational velocity ($v_{\rm c}$) is the value predicted from the enclosed mass at 0.5 \rgalStop. Full rotation curves for most galaxies in our sample can be found in \citet{hopkins20}. $M_{*}$ is the stellar mass contained within 3 $R_{*,1/2}$.}
\end{table*}

This study is based on 4 simulated Milky Way-mass disky galaxies evolved in cosmological context (see Fig.~\ref{fig:ba_dens} for face-on and edge-on view of their gaseous disks) where gas infall, large and small scale outflows and galaxy interactions are modeled self-consistently. 
Simulations utilize "zoom-in" technique to reach high resolution in fully cosmological settings and were run with the gravity+(magneto)hydrodynamics code GIZMO \citep{hopkins15} using mesh-free Lagrangian Godunov (meshless finite mass, MFM) method.

Cooling, star formation and stellar feedback are implemented as in FIRE-2 \citep{hopkins18}, an updated version of the FIRE project \citep{hopkins14}. The simulations include photoionization and photoheating from the cosmic UV background based on the \cite{faucher09} model.
Star formation is enabled in self-shielding/molecular, Jean unstable dense ($\rm n_{\rm H} > 1000 \rm cm^{-3}$) gas. 
Once created, star particles are treated as single-age stellar populations with IMF-averaged feedback properties calculated from STARBURST99 \citep{leitherer99} assuming a \cite{kroupa01} IMF. Feedback from SNe (Type Ia and II), stellar mass loss (O/B and AGB), and radiation (photo-ionization, photo-electric heating, and UV/optical/IR radiation pressure) are explicitly treated as in \citet{hopkins18}. 
In this analysis we use simulations that in addition to standard FIRE-2 feedback incorporate magneto-hydrodynamics (MHD) and cosmic ray (CR) physics \citep{hopkins20} using CR transport methodology fully described in \cite{chan19}. To summarize, runs include CR injection in SNe shocks, fully-anisotropic CR transport with streaming, advection, and diffusion, CR losses (hadronic and Coulomb, adiabatic, streaming), and CR-gas coupling.

 The FIRE simulations have been successful in matching a range of galactic properties to observations, including total stellar mass; star formation rates and histories \citep{hopkins14,sparre17,feldmann16,santistevan20}; dense HI covering fraction in the circumgalatic medium at both low and high redshift \citep{faucher15,faucher16,hafen17}; outflow properties \citep{muratov15, muratov17}; metallicities \citep{ma16a,bellardini21}; morphological/kinematic structure of thin/thick disks \citep{ma17b,yu21,sanderson20}; baryonic and dark matter mass profiles and content within the halo \citep{chan15, wetzel16}; Giant molecular cloud properties \citep{benincasa20,guszejnov20}; and circular velocity profiles \citep{hopkins18}.

 In general, our MW-mass galaxies with CRs show good agreement with the observationally inferred M$^{*}$-M$_{\rm{halo}}$ relation and are disk dominated \citep{hopkins20}. Stellar masses in runs without CRs are somewhat higher than this relation predicts. On the other hand, CR+ runs may be slightly underestimating stellar mass, although statistics are still poor.
 Our choice of simulations with additional CR physics is mainly guided by their lower late time star formation rates of 2-3 $\rm M_{\odot}/yr$ vs 3-10 $\rm M_{\odot}/yr$ in FIRE-2 simulations without CRs, which is closer to what is seen in the Milky Way.  This lower star formation is associated with lower velocity dispersion of galactic disks and potentially additional planar alignment of the accreting gas \citep{hopkins21_crOutflows}. In addition, the circum-galactic medium in MW-mass simulations with CRs is in better agreement with observations of low and intermediate ions seen in absorption systems around galaxies \citep{ji20}. We compared the results from our default CR+ runs with simulations without CRs and found that, despite differences in late-time star formation, both default FIRE-2 simulations and CR+ runs show similar {\it qualitative} behavior and trends in gas accretion onto low redshift disks (see Appendix~\ref{sec:appendix_comparisons}).

The minimum baryonic particle mass in our simulations is $\rm m_{b,min} = 7100 M_{\odot}$ and a typical gravitational force softening for star forming gas is $\sim 2$ pc. Note that spatial resolution (softening and smoothing lengths) for our gas particles is adaptive; typical force softening for ISM gas is $\sim 20$ pc. All simulations employ a standard flat $\Lambda$CDM cosmology with $\rm h \approx 0.7$, $\rm \Omega_{M} = 1-\Omega_{\Lambda} \approx 0.27$, and $\rm \Omega_{b} \approx 0.046$ (consistent with \cite{planck14}).

\section{Simulation Analysis}\label{sec:analysis}

In Fig.~\ref{fig:ba_coldens} we show visualizations of the HI and H$_{2}$ column densities viewed face and edge for our four simulated galaxies; a basic summary of their relevant properties is presented in Table~\ref{table:disks}. We define the radius of each galaxy as the radius where the face-on, azimuthally averaged neutral hydrogen column density drops below $10^{20.3} \rm{cm}^{-2}$ (\rgalStop), the column density limit of Damped Lyman Alpha (DLA) systems \citep{wolfe05}. We additionally provide the radii at which this column density drops below $10^{19} \rm{cm}^{-2}$. The disk scale height is quantified as the height the average gas density drops by a factor of \textit{e} from the average value within $\pm$20 pc of the midplane. It is quantified in two ways, based on total hydrogen number density ($h_{\rm total}$) as well as cold (T $<$ 100 K) hydrogen number density  within the inner 5 kpc ($h_{\rm cold,inner}$). We additionally calculate a characteristic time period for each galaxy (\tgal) as the rotation period of gas particles at the disk edge. Rotational velocity ($v_{\rm c}$) was predicted from the enclosed mass at 0.5 \rgalStop. Full rotation curves for most of these galaxies can be found in \cite{hopkins20}. These values will be used to normalize distances and times in subsequent sections to allow for more direct comparisons between galaxies.

 We focus on the gas in late time disks in the redshift interval z=0.0-0.2, as the disks are thin, stable, and have a clear orientation. For each galaxy, we initially define a cylindrical coordinate system centered on the galactic center and oriented along the angular momentum vector of the galaxy. We calculate the galactic center from the mass distribution of the star particles using a shrinking sphere algorithm.\footnote{A 1 Mpc radius sphere was defined around the region of maximal gas density and the stellar center of mass was calculated. This sphere was shrunk by a factor of 0.7 and the center of mass recalculated until it reached a radius of 10 kpc. The centering was repeated without shrinking until the center converged to a stable value.} We determine the galactic velocity by calculating the mass weighted velocity average of all star and dark matter particles within 15 kpc of galactic center. The orientation of the angular momentum vector was determined from the vector sum of angular momenta of cold dense gas particles (T $<$ 8000 K, $n > 1 \rm cm^{-3}$ ) in the inner 10 kpc, with respect to the galactic center. We analyze the gas flow properties through a bulk flow analysis of all gas particles, as well as particle tracking of sub-selections of accreting particles.

\subsection{Bulk Flow Analysis}\label{sec:analysis-ba}
The mass, density, velocity, and position of each gas particle were read from each simulation snapshot. We transformed the position and velocity vectors into the previously defined cylindrical coordinate system and binned the data in order to calculate locally averaged values. Data were binned into a 0.25 $\times$ 0.25 $\times$ 0.25 kpc Cartesian grid (x,y,z) and into a 0.1 kpc $\times$ 0.1$\pi$ rad $\times$ 0.25 kpc cylindrical grid (r,$\theta$,z). In order to quantify global gas flow properties, we additionally binned data in 1 kpc spherical radial bins. During analysis, bins were averaged or summed together to create coarser resolution as needed. For each bin, we calculated the total mass, angular momentum, and linear momentum components. We additionally calculated mass weighted average velocity components and specific angular momenta. We estimated mass flux as follows:

\begin{equation}
    \rm{Mass} \: \rm{Flux} = \frac{1}{\Delta \, r} \sum_{i}  \vv{p_{i}} \cdot \hat{r}_{i}
\end{equation}

where $\Delta \, r$ is the size of the radial bin and $\vv{p_{i}} \cdot \hat{r}_{i}$ is the i$^{\rm th}$ particles momentum component in the radial direction (spherical or cylindrical).

For some of our visualizations and observational predictions we utilize the column densities of total hydrogen, HI, and H$_2$. Neutral hydrogen fraction is calculated using on-the-fly self shielding in FIRE with a local Sobolev/Jeans-length approximation \citep{hopkins18} and is written in the simulation output. We additionally take molecular fraction values directly from the simulations, which were calculated during the simulation runs following the approximation outlined in \citep{krumholz11}. \footnote{FIRE-2 simulations by do not save molecular fraction as an output by default. We therefore reran the simulations at redshift 0 soley to output molecular fraction. The timestep was chosen such that no particles were updated.}

Stellar mass distribution and SFR were binned spatially in a similar fashion. For each stellar bin, we calculated the total stellar mass by summing the mass of each star particle in each bin. In order to quantify star formation, we calculated the total mass of newly formed stars, identified as stars formed within 20 Myrs. This short time interval minimizes the effects of stellar drift in identifying star formation location \citep{orr18}.

\subsection{Particle Tracking}\label{sec:analysis-pt}
Resolution elements (particles) in the FIRE-2 simulations are uniquely trackable throughout a run, including through particle splitting and star formation events.\footnote{Each particle is assigned a particle ID, child ID, and generation number. When a gas particle crosses a certain mass threshold, it is split into two separate particles with unique child IDs and the particles generation number is incremented. Star particles retain the original gas particles IDs.} To characterize the behavior of individual gas elements as they join the disk and form stars, we defined accreting gas as particles that are present inside the characteristic radius of the disk (\rgalStop) at z=0.03 (either as gas or stars), but are located at a distance greater than this radius at redshift z=0.17. We track particles between redshift z=0.20 and z=0.00.The difference from the selection criteria (z=0.03-0.17) is such that we follow each particle for at least 0.4 Gyr before and after accretion.\footnote{If multiple particles split from the same parent during this time period, the parent particle is accounted for multiple times. This double counting is corrected for when calculating relevant metrics.}

At every snapshot, we track each particles' position, momentum, mass, and particle type (gas or star) in order to characterize the kinematics of the accretion process and identify where/when they form stars. We identify when/where a particle joins the disk in 2 ways: (i) geometrically and (ii) by specific angular momentum. In the geometric classification, a particle joins the disk when it passes through a height of 2\hgal\ and remains for at least one \tgal. In the specific angular momentum classification, a particle joins the disk when its specific angular momentum reaches within 20\% of the rotation curve as predicted by the mass distribution and remains for at least one \tgal. Both of these cutoffs are based on the typical spread of particle heights and angular momenta within the disk. Reducing the time requirement from one \tgal causes particles to be classified as joining the disk at larger radii on average.

\section{Results} \label{sec:results}

\begin{table}
\centering
\begin{tabular}{p{1.75cm} p{1.75cm} p{1.75cm} p{1.75cm}   }
\hline
\hline
Galaxy & Total Accretion & Disk Mass Flux & SFR \\
& [M$_{\odot}$/yr] & [M$_{\odot}$/yr] & [M$_{\odot}$/yr] \\
\hline
\bf{m12m} & 1.7 $\pm$ 0.5 & 2.7 $\pm$ 0.4 & 2.5 $\pm$ 0.6 \\

\bf{m12i} & 2.2 $\pm$ 1.0 & 1.5 $\pm$ 0.3 & 1.7 $\pm$ 1.5 \\

\bf{m12f} & 3.7 $\pm$ 10.3 & 4.4 $\pm$ 1.6 & 2.3 $\pm$ 2.0 \\

\bf{m12b} & 2.9 $\pm$ 1.9 & 1.8 $\pm$ 0.7 & 1.9 $\pm$ 1.3 \\
\hline
\hline

\end{tabular}

\caption{\label{table:ba-mf} Comparison between total accretion, mass flux through the disk, and SFR within the disk, averaged from redshift z = 0.1 to 0. Total accretion was calculated from the spherical radial mass flux through a 1 kpc spherical shell at 1.5 \rgal (Table~\ref{table:disks}). Disk Mass Flux was calculated from the cylindrical radial mass flux through a cylindrical shell at 0.5 \rgal and limited to 2 times the disk scale height. SFR was calculated from the total SFR interior to 1.5 \rgalStop.}
\end{table}

All four galaxies in our sample (\textbf{m12m}, \textbf{m12i}, \textbf{m12f}, and \textbf{m12b}) have large disky structures. In general $\rm H_{2}$ is centrally concentrated and correlated with spiral arm-like structure, while HI is more  smoothly distributed throughout the disks. Typical face-on HI column densities in the disk region are of order $10^{21} \rm cm^{-2}$ and drop off very sharply at the disk edge (Fig.~\ref{fig:ba_coldens}). There are some key structural differences of note between galaxies. The disk in {\bf m12m} is much more extended than the other galaxies, while the {\bf m12b} is the most compact. This difference affects location and geometry of the gas accretion and star formation. Additionally, \textbf{m12f} has a minor merger at z$\sim$0.1 with an LMC mass object (M$_{*}$=1.7$\times$10$^{9}$ M$_{\odot}$).

\subsection{Global Gas Flow Properties}\label{sec:results-global-flow}

\begin{figure} 

	 \center{\includegraphics[width= .48 \textwidth]
	    {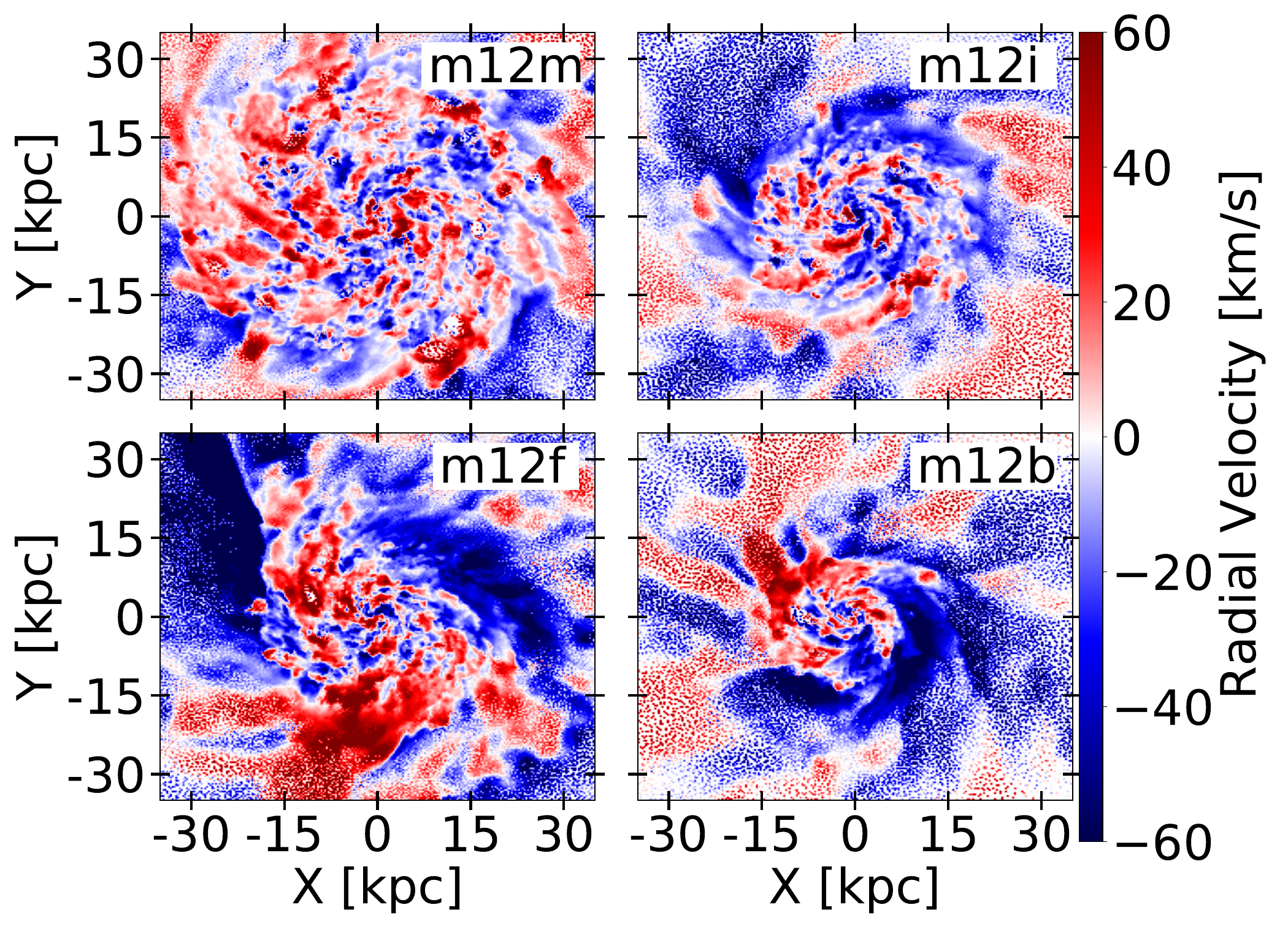}}
	  \caption{\label{fig:radvel-maps} Face on views showing averaged cylindrical radial velocities for all gas within 1.5 kpc of the disk plane at z=0. For a corresponding figure with vertical velocity, see Appendix~\ref{sec:appendix_II}. Note the fast inflowing gas in the top right quadrant of \textbf{m12f} corresponding to continued accretion from the merger event. All galaxies show clear inflow at radii outside the disk.
	  }
\end{figure}
	
\begin{figure*}

	  \center{\includegraphics[width= \textwidth]
	       {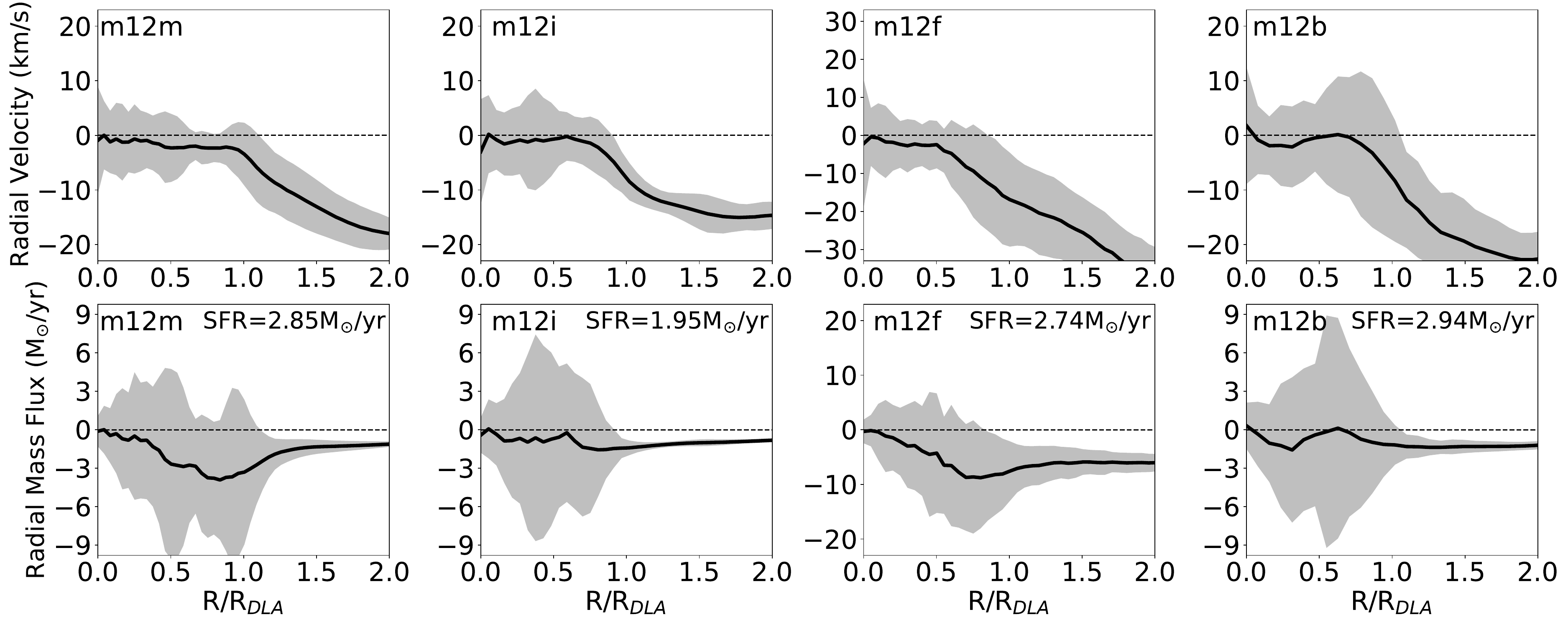}}
	  \caption{\label{fig:ba-radvel_and_radmf}
	  Cylindrical radial velocity and mass flux as a function of cylindrical radius measured in 1 kpc radial bins and averaged over 1 \tgal (Table~\ref{table:disks}). Values were averaged between +/- 10 kpc from the disk plane. Changing this height cutoff has minimal effects on the curves within the disk, but smaller cutoffs significantly reduce the mass flux curve outside the disk. Shaded regions represent the standard deviation of values over time. Bin-averaged radial velocity is $\sim 10 \rm km/s$ outside of the disk, but drops to low values within the disk, except in the very center, which shows higher velocities in some galaxies. Radial mass flux increases going inwards toward the disk, peaking slightly interior to the disk edge, although \textbf{m12b} deviates from this trend, largely due to it's smaller disk size and more centrally concentrated star formation. This peak flux is typically of order of galactic SFR (shown in top right corner for reference), except in {\bf m12f} where a recent merger is driving a strong increase in gas accretion.
	  }
\end{figure*}

All galaxies show net inflow at larger radii (Fig.~\ref{fig:radvel-maps}, \ref{fig:ba-radvel_and_radmf}). The mass flux from this inflow is roughly equivalent to both the radial mass flux through the disk, as well as the total star formation rate of the galaxy (Table~\ref{table:ba-mf}). Values are not expected to match perfectly owing to low level galactic outflows \citep{muratov15, pandya21} and partial gas supply from stellar mass recycling within the disk. Fig.~\ref{fig:radvel-maps} shows face on maps of cylindrical radial velocity. While there are outward flowing features throughout, the disk outskirts are inflow dominated with significant azimuthal velocity variation. Fig.~\ref{fig:ba-radvel_and_radmf} quantifies the cylindrical radial velocity and radial mass flux as a function of radius. Within the disk, average radial velocities are on the order of a few km/s. Within 100 Myr of z=0, the average radial velocities through the disk from \textbf{m12m}, \textbf{m12i}, \textbf{m12f}, and \textbf{m12b} are -1.2, -0.4, -4.2, and -1.5 km/s, respectively.

Inflowing gas outside of the disk rapidly slows down by around 10 km/s as it interacts with disk material. The shaded regions in Fig.~\ref{fig:ba-radvel_and_radmf} show the standard deviation over the time period analyzed. The spikes correspond to the pattern of inflow and outflow seen in Fig.~\ref{fig:radvel-maps} and are likely related to spiral arm-like structure. 

The radial mass flux curve peaks slightly interior to the shift in radial velocity, possibly by momentum transfer from the infalling gas that joins the disk. Mass flux values decrease gradually with radius outside the disk, and rapidly approach zero towards the center of the disk as particles become locked up in stars and some are removed out of the disk in a wind/fountain, typically in the vertical direction \citep{chan_inPrep}. On average, there is still net radial inflow at all radii, fueling disk growth.

\subsection{Angular Momentum}\label{sec:results-angular-momentum}

\begin{figure*}
	  \center{\includegraphics[width= \textwidth]
	       {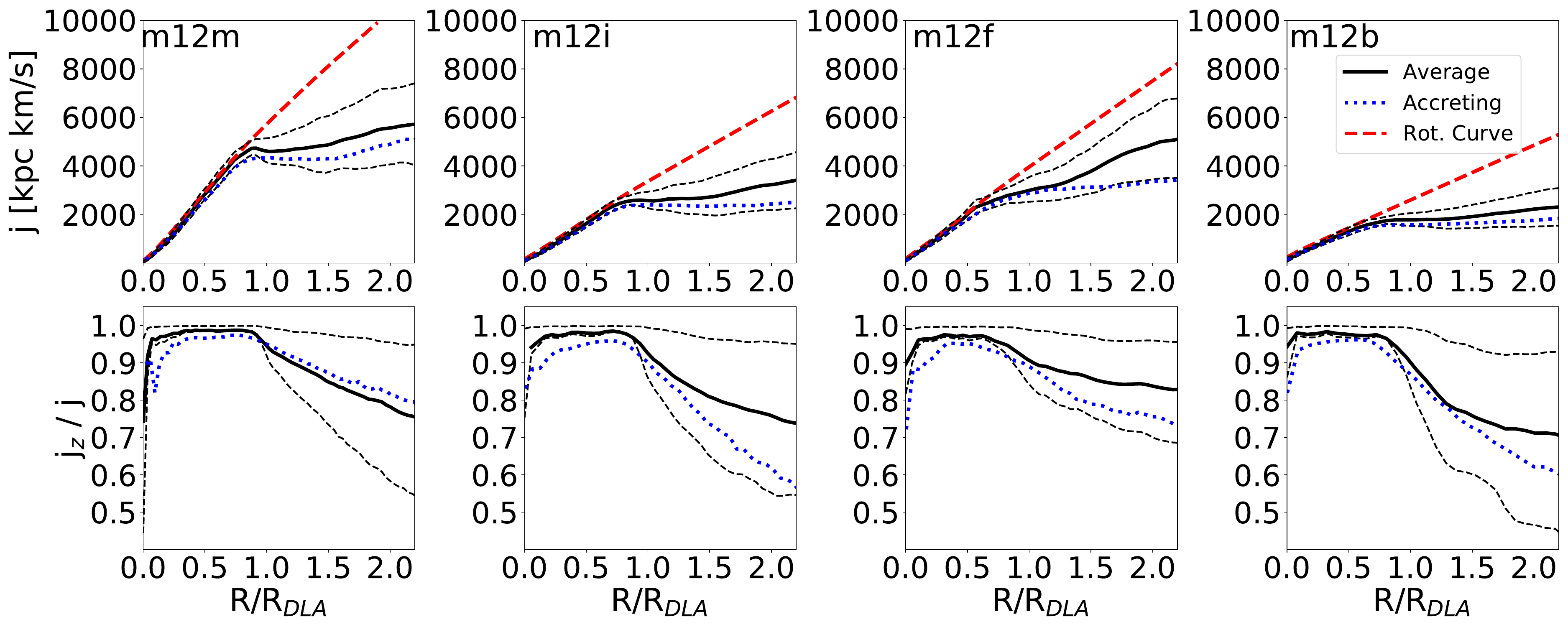}}
	  
	  \caption{\label{fig:ba-j-graphs} \textit{Top Row:} Specific angular momentum as a function of distance from galactic center at z=0. The thick-dashed red line represents the expected rotation curve inferred from the mass distribution. The black dashed lines show the standard deviations. The blue dashed lines are the mass-weighted average specific angular momenta of gas elements from the sample of accreting particles described in Section~\ref{sec:analysis-pt} 1 \tgal prior to accretion. Note that $j$ increases very slowly with radius in the CGM (beyond $R_{\rm DLA}$), and remains only slightly elevated from $j$ at the disk edge. \textit{Bottom Row:} Average z-component of $\vec{j}$ normalized to the total specific angular momentum of each particle. This is a metric of how well particles' rotation is aligned with the disk. More compact disks (e.g. \textbf{m12i}, \textbf{m12b}) show a sharper drop off at the disk edge, though all gas is still strongly co-rotating at these scales. Graphs are generated with radial bin resolution of 1 kpc.
	  }
\end{figure*}

Fig.~\ref{fig:ba-j-graphs} shows specific angular momentum of gas ($\vec{j}$) as a function of galacto-centric distance. Within the disk, this curve follows the rotation curve of the galaxy predicted by the enclosed mass very closely.\footnote{We use mass enclosed in a sphere of a given radius and for simplicity assume spherical mass distribution when calculating $\rm v_c$, which is sufficiently accurate for our analysis here.} For $\rm r \leq R_{\rm DLA}$, the specific angular momentum increases with radius roughly linearly as expected for a flat rotation curve. The magnitude of $\vec{j}$ flattens at r $>$ \rgal with only a slight increase with radius as it falls below the value needed for rotational support (Fig.~\ref{fig:ba-j-graphs}, top). Note that gas in the inner CGM still has significantly higher specific angular momentum then the gas within the disk on average, consistent with $|\vec{j}|$ trends found in the CGM of FIRE galaxies in \citet{elbadry18}.  The spread of $|\vec{j}|$ outside the disk increases in all four galaxies. The average $|\vec{j}|$ of gas accreting within 2 \tgal (blue dashed line in Fig.~\ref{fig:ba-j-graphs}) is slightly lower than the overall gas average, especially at larger radii.\footnote{Note that here we mark a gas particle as accreted if it passes within 2 scale-heights of the disk (see Section~\ref{sec:results-joining-the-disk}).}

While the gas at  $r >$ \rgal is not fully rotationally supported, its motion is still strongly aligned with the rotation of the galaxy. The spread in the angular momentum alignment increases for $r > R_{\rm DLA}$ (Fig.~\ref{fig:ba-j-graphs}, bottom). Accreting particles are on average well aligned with the disk but slightly less so than the general inner CGM. A larger degree of misalignment in the central regions is caused by a small fraction of particles that accrete at larger angles, while most join the disk at small angles in the outskirts (Section~\ref{sec:results-joining-the-disk}).

Individual gas mass elements accrete onto a galaxy while approximately conserving angular momentum from 2-3 \rgal until they reach corotation with the galaxy at $\lesssim$ \rgalStop. In detail, they do lose a small amount of angular momentum as they accrete but these losses are very small, with $dj/dr \sim$ 10-30 kpc km/s /kpc on average (slightly higher in {\bf m12f}). This is similar to what is shown in \cite{danovich12,Danovich15}, which shows that for high-z galaxies, inflowing gas in the halo does not lose a significant amount of angular momentum until it is near the disk.

The specific angular momentum of the gas outside the disk has an effect on where it will reach full rotational support. Gas with $j>j_{\rm disk-edge}$ will lose the small excess specific angular and settle near the disk edge. Gas with $j < j_{\rm disk-edge}$ will typically still lose a small amount of angular momentum and enter corotation at a more interior radius.\footnote{In rare cases, accreting gas mass element can have abnormally high specific angular momentum and can lose angular momentum at a rate up to 200 kpc km/s / kpc as it is torqued into corotation.} Once within the disk, gas tends to follow the rotation curve closely. During slow radial flow inward gas remains in corotation, resulting in average specific angular momentum loss $dj/dr \sim 200$ kpc km/s /kpc.

\subsection{Joining the Disk}\label{sec:results-joining-the-disk}

 \begin{figure}
	  \center{\includegraphics[width=.48\textwidth]
	       {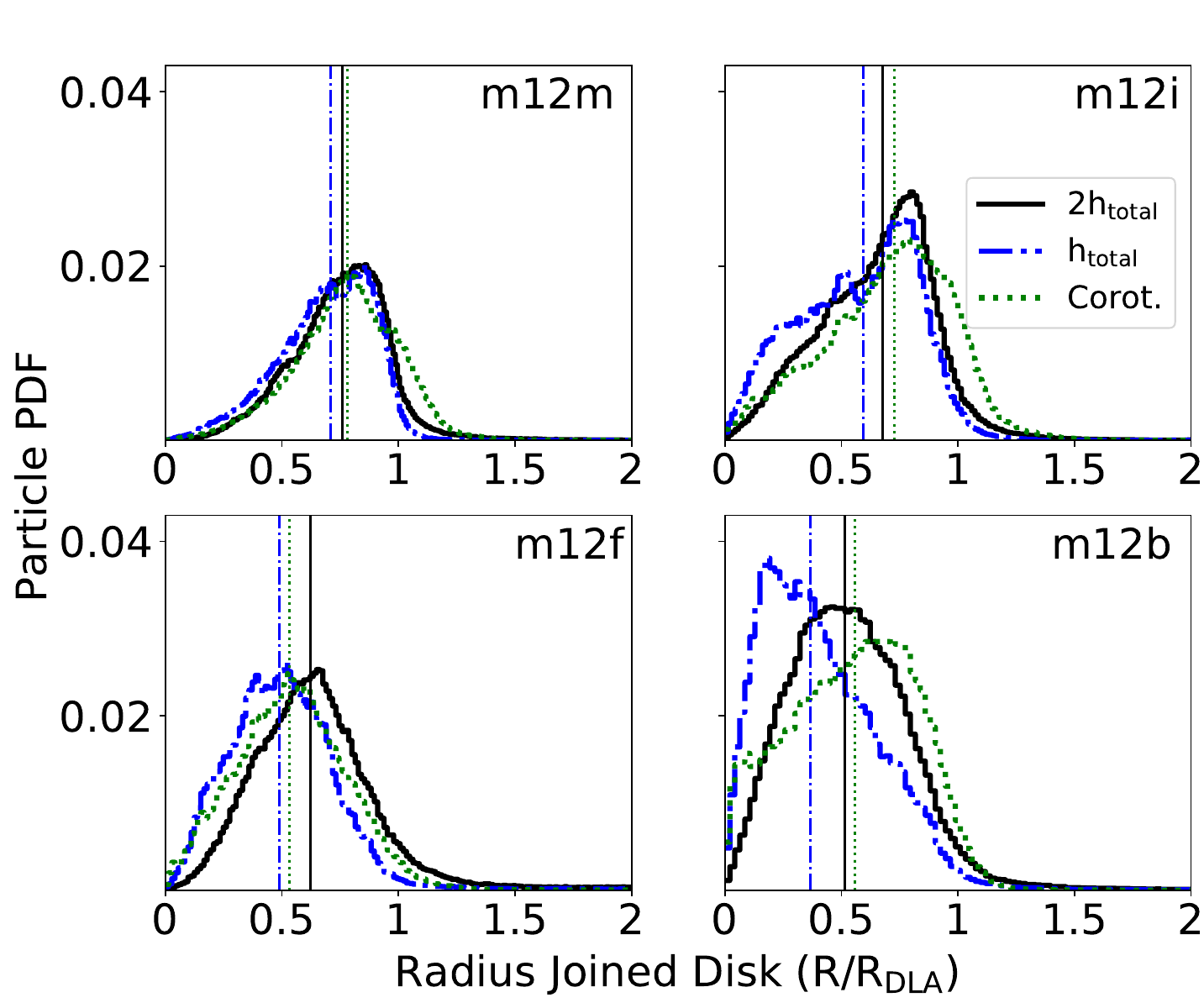}}

	  \caption{	 \label{fig:pt-join}
	  Radius at which a particle (from the population tracked from z=0.2 to z=0) joins the galactic disk and remains for at least \tgal (Table~\ref{table:disks}). The black (blue dashed) histogram shows where particles pass 2 scale heights (1 scale height) of the disk. The green dotted histogram shows particle's radial position when its specific angular momentum is within 20\% of the value needed for rotational support and remains such for 1 \tgal. Histograms are weighted by accreted gas mass. Vertical lines show median values. Particles preferentially join the disk slightly interior to the disk edge ($R_{\rm DLA}$, Table~\ref{table:disks}), with most particles joining outside of 0.5 \rgalStop. See Fig.~\ref{fig:pt-rJoin-areaWeight} for accretion per unit area.
} 
\end{figure}

 In Fig.~\ref{fig:pt-join} we quantify the radius at which gas joins the disk based on where particles enter 1 or 2 times the scale-height of the disk and remain for 1 \tgal, as well as where their specific angular momentum reaches within 20\% of the rotation curve and remain for 1 \tgal (Section~\ref{sec:analysis-pt}). Particles tend to enter co-rotation with the disk at roughly the same radii at which they join the disk geometrically, although there are differences between galaxies.
Particles predominantly join the disk within a few kpc of the disk edge (\rgalStop), which is close to \rgal for {\bf m12m} and {\bf m12i}, but closer to 0.5\rgal for the more compact disk in {\bf m12b}. When looking at the medians of the distributions, most particles join the disk outside of 0.5 \rgal, with the exception of the 1 scale-height definition for \textbf{m12b}. We note that \textbf{m12b} is an interesting case compared to the larger, more well defined disks, showing gas accreting more orthogonally, joining closer to the center, and showing more centrally concentrated star formation. Relaxing the time criteria pushes this distribution to larger radii, slightly closer to the disk edge. In all cases this is well outside of the bulk of the stellar component of the galaxy, typically at $\sim 3\times R_{\rm *, 1/2}$.
Our default criteria for when/where a mass element joins the disk will be the geometric classification at two scale heights (black, solid histogram).

Note that when expressed as accretion rate per unit area (see Figure \ref{fig:pt-rJoin-areaWeight}), these distributions tend to be flatter and are in  broad agreement semi-analytic models of galactic disk evolution that require such flat distributions to match a broad range of disk scaling relations \citep{forbes19}. However, in the most extended disk in our sample (\textbf{m12m}), even the accretion per area is clearly peaked near the disk edge.

\begin{figure*}
	  \center{\includegraphics[width= \textwidth]
	       {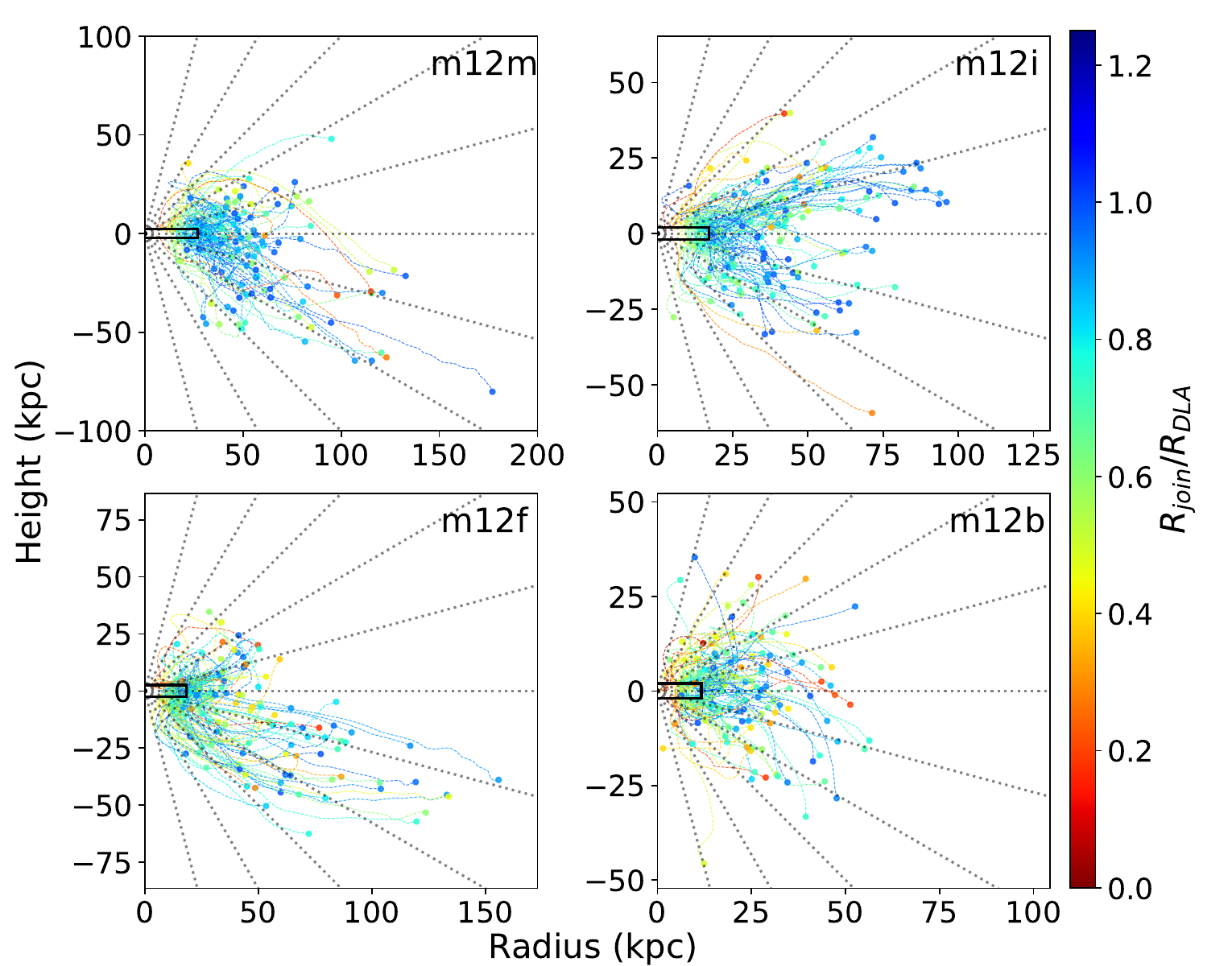}}
	  \caption{\label{fig:pt-trajectories}
      Lines visualizing the trajectories for a set of 250 randomly sampled particles from the distribution in Fig.~\ref{fig:pt-angles}. The galactic disk is visualized by the solid black lines representing extent of \rgal and thickness of $\pm$h$_{\rm total}$. The dashed grey lines show increments of 15$^{\circ}$. Trajectories are shown from when the gas elements join the disk to 1.5 \tgal for visualization purposes. Lines are color coded by the radius at which they join (redder lines correspond to particles that join more interior, colorscale saturates at 1.25 times the radius of the disk). Particles that join more interior tend to have trajectories at higher polar angles. Funnel structure is clearly visible for accreting particles.
      }
\end{figure*}

\begin{figure}
	  \center{\includegraphics[width= .5 \textwidth]
	       {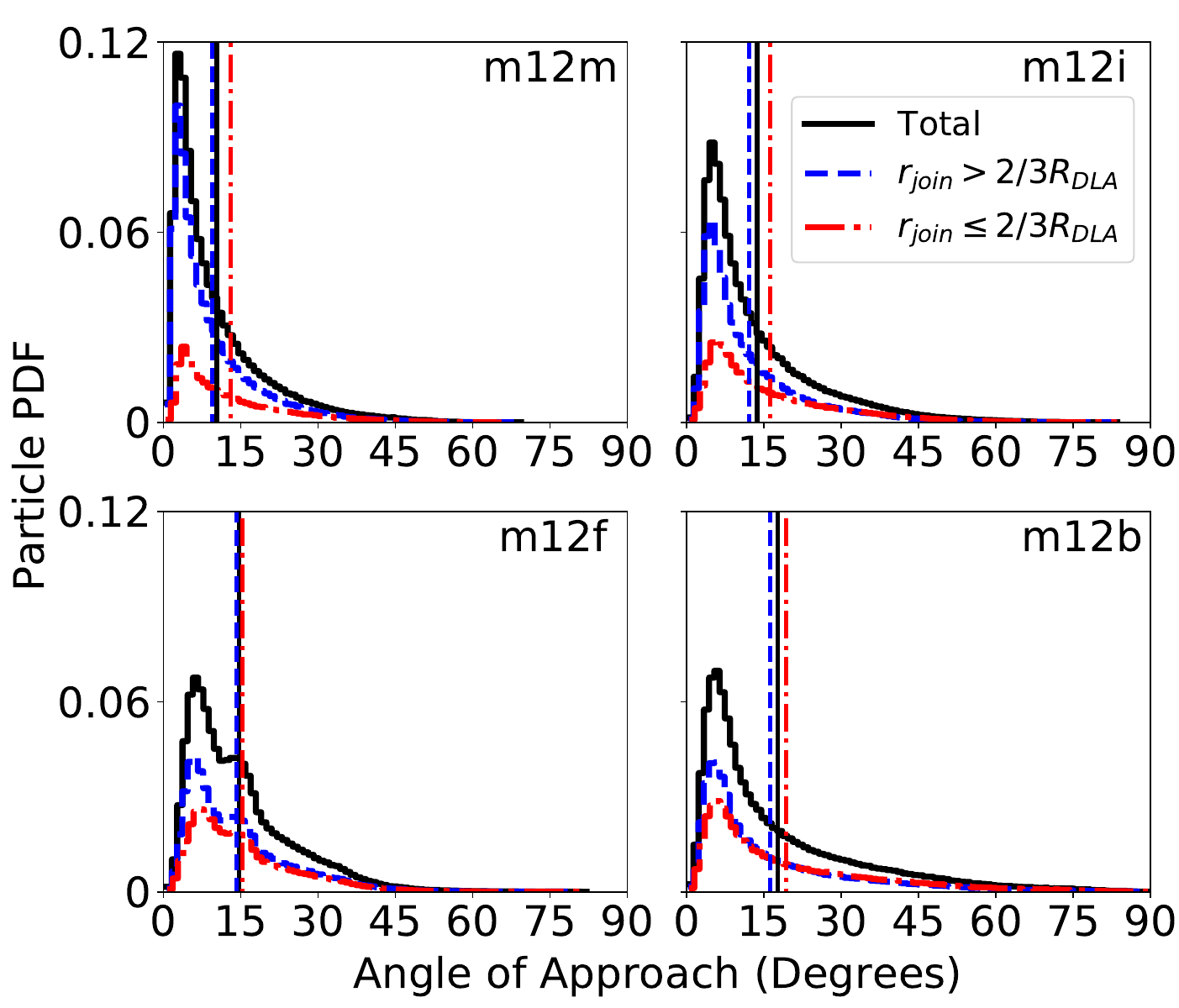}}
	  \caption{\label{fig:pt-angles}
	 Angle above/below the disk plane with which particles accrete on the galaxy, i.e. pass within twice the scale height of the disk and remain for at least 1 \tgal. Angle is averaged over the trajectory of the particle for 1 \tgal prior to joining the disk and is measured from galactic center. The centroid of each distribution is displayed as the vertical dashed line. Colors of the histograms are based on where the particles join the disk. Most particles join the disk at non-zero angles within 15 degrees. Particles that join the disk more centrally tend to accrete at larger angles.
     }
\end{figure}

\begin{figure}
	  \center{\includegraphics[width= .5 \textwidth]
	       {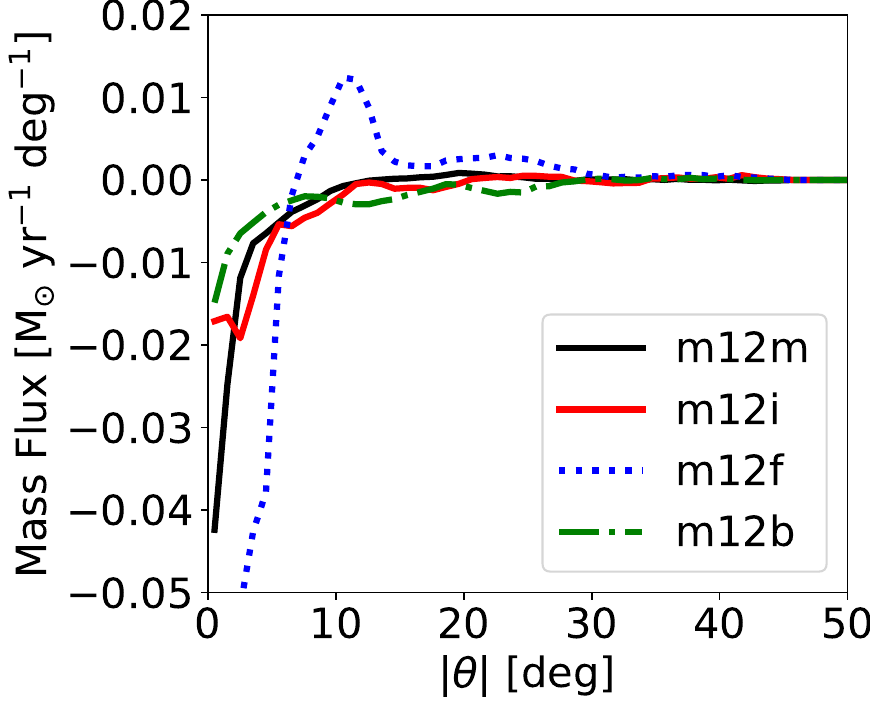}}
	  \caption{\label{fig:ba_MF_v_PolarAngle}
	 Radial mass flux per degree (averaged over 1 \tgal) as a function of angle above/below the disk plane. Average angle of inflow is around 8 degrees. Mass flux was calculated through a 1 kpc spherical radial bin centered at 1 \rgal with 2 degree polar angle bins. Note that \textbf{m12f} shows outflowing gas at larger angles owing to the recent merger event.
     }
\end{figure}

Fig.~\ref{fig:pt-trajectories} and Fig.~\ref{fig:pt-angles} characterize the trajectories individual particles undergo as they accrete onto the disk. As seen in Fig.~\ref{fig:pt-trajectories}, during accretion particles tend to move parallel to the galactic plane at large distances, before falling onto the disk. The average polar angle (as measured from the center of the galaxy) of this accretion process is therefore non-zero but small (Fig.~\ref{fig:pt-angles}), with typical values around 15$^{\circ}$ as measured from galatic center. This corresponds to flare or funnel like structures near the disk edge. The angle at which these particles accrete is correlated with where they end up joining the disk. Particles accreting at higher polar angles tend to join the disk at smaller radii.

The angle of these trajectories tends to steepen over time, resulting in larger angles of approach near the location where gas joins the  disk. Particles approaching at higher angles ultimately "fall" onto the disk just prior to joining, showing larger vertical velocities. This effect is still present on average in particles joining closer to the edge, but is  much milder. The overall trajectories of accreting gas prior to and after they join the disk are still predominantly parallel to the disk. We emphasize that Fig.~\ref{fig:pt-angles} shows the mean angle over a single \tgal, which will be smaller, on average, than the angle just prior to accretion event. In addition, gas accreting near the plane of the disk tends to have lower radial velocities. In Fig.~\ref{fig:pt-trajectories}, these particles will be more clustered near the disk edge, making it difficult to distinguish individual trajectories and enabling gas at larger angles to have relatively stronger visual impression in this figure.

To show our selection of accreting particles in Fig.~\ref{fig:pt-trajectories} and Fig.~\ref{fig:pt-angles} roughly corresponds to what is seen in the total mass flux, we look at the polar angle in terms of radial mass flux of all gas binned in a 1 kpc spherical shell at 1 \rgalStop. We find that the average angle is slightly smaller ($\sim$ 8 degrees), but is still small and nonzero (Fig.~\ref{fig:ba_MF_v_PolarAngle}).

\subsection{Directional Rates}\label{sec:results-directional-rates}
	
\begin{figure}
	  \center{\includegraphics[width=.5 \textwidth]
 	       {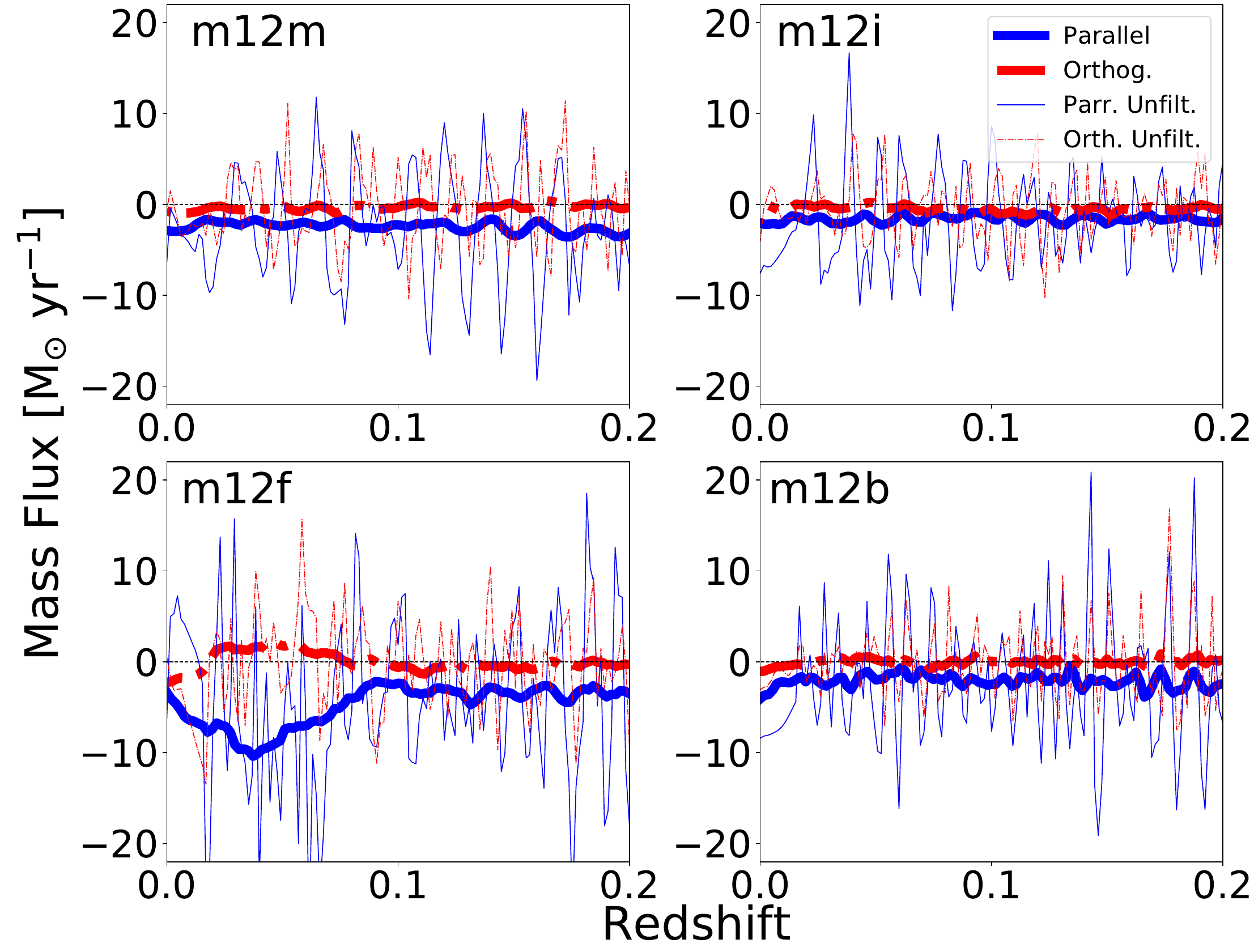}}
	  \caption{\label{fig:MF-directional-quant} Mass fluxes within the disk. Blue, solid lines show mass fluxes parallel to the disk. Red, dashed lines show mass fluxes orthogonal to galactic disks. Negative values represent in-flowing material. Parallel mass fluxes (blue) are measured within 1 scale height of the disk through a 0.5 kpc radial bin centered at 0.5 \rgalStop. Orthogonal mass fluxes are measured within 0.5 \rgal in a 0.5 kpc vertical bin at 1 scale height above the disk. Both scale heights  and \rgal were calculated at each time point as described in Table~\ref{table:disks} to account for changes in disk size. The thick lines are smoothed using a moving average filter with a window width equal to the dynamical time for each galaxy in order to suppress oscillations. Unfiltered values are shown by thin lines. Note the amplitudes of the smoothed parallel mass fluxes tend to by higher and are consistently inflowing, while the orthogonal mass fluxes show inflow and outflow periods. 
}
\end{figure}

On galactic scales, gas accretes into star forming regions largely parallel to the disk plane, as opposed to vertically. Fig.~\ref{fig:MF-directional-quant} shows that the time averaged vertical mass flux is around an order of magnitude lower than the parallel component. Galaxies change between periods of net vertical inflow and outflow, while net radial mass fluxes are consistently inflowing. Vertical inflow can still be significant for some short time periods. For instance, the merger event in \textbf{m12f} leads to a period where vertical inflow is as almost as significant as radial inflow at redshift 0. We note that parallel infall in Fig.~\ref{fig:MF-directional-quant} measured at 0.5 \rgal is largely from the gas co-rotating in the outer disk that moves radially inward as the accretion from the CGM has largely settled into the disk further out. If we constrain our measure of vertical flows to the innermost regions of the galaxy only, we see more consistent vertical outflows. These features ultimately arise from stellar feedback, and typically only have values of 0-0.5 \MFunit.

\begin{figure}
	  \center{\includegraphics[width=.45 \textwidth]
	       {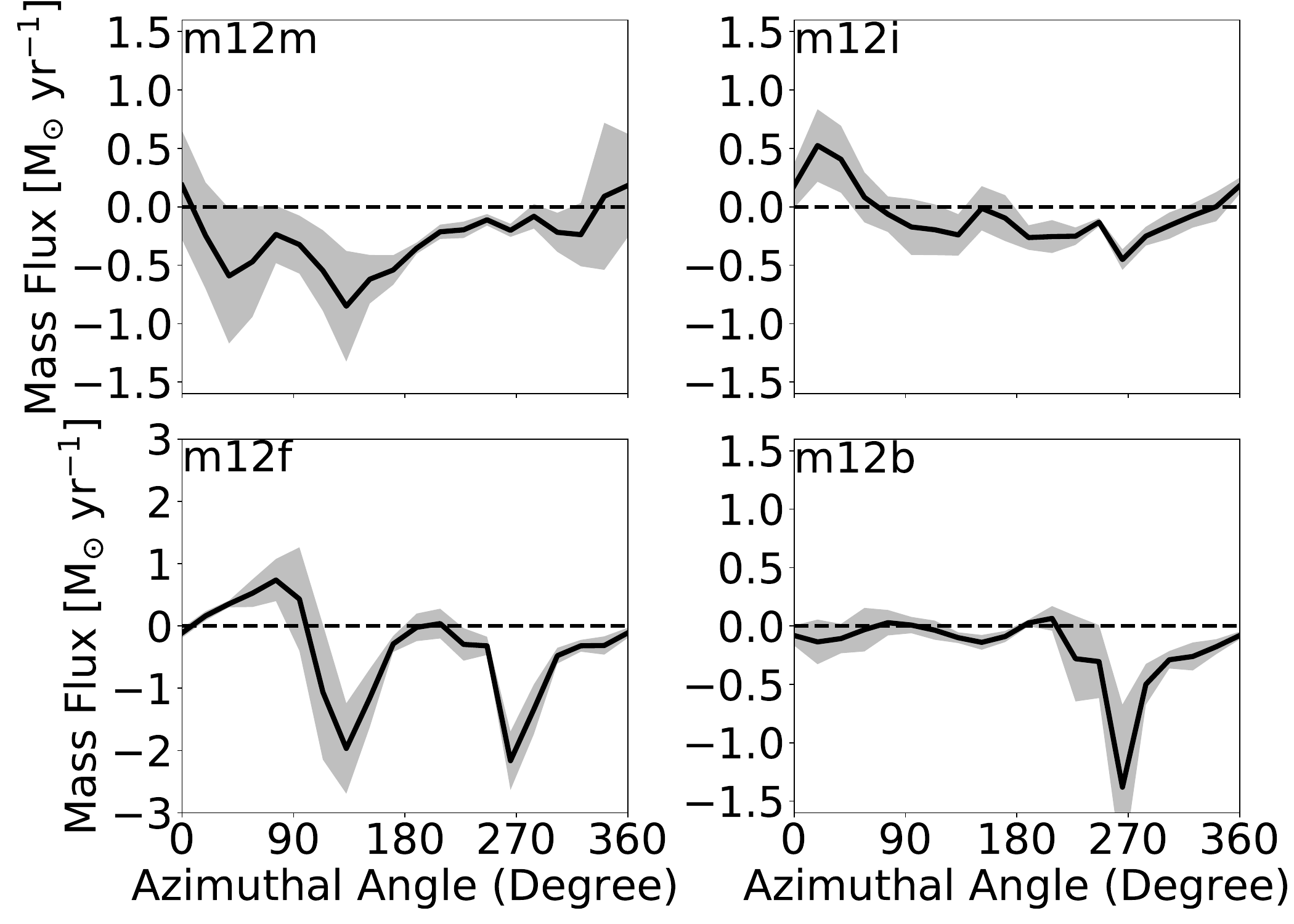}}
	  \caption{\label{fig:ba_azimuthaldist} Cylindrical radial mass flux through a 1 kpc radial bin at the disk edge (R$_{\rm DLA}$, Table~\ref{table:disks}) as a function of azimuthal angle on the disk. Data was averaged over 0.25 \tgal near redshift z=0 for all particles within +/- 2 $h_{\rm disk}$ of the disk plane. Standard deviation over time is shown by the shaded region. Dips in mass flux come largely come from radial oscillations of both the disk material near the edge as well as radial oscillations of the accreting material. Anisotropic distribution of accreting gas also plays a role depending on the galaxy (e.g., the merger event in \textbf{m12f}).  Figure is averaged over a relatively short period to emphasize angular  differences at a given time.
}
\end{figure}

This trend can be seen on local scales in Fig.~\ref{fig:radvel-maps} and Fig.~\ref{fig:vertical_velocity_maps}, which show a color map corresponding to radial and vertical velocity components, respectively. Gas accretes asymmetrically at the disk edge, with each galaxy showing regions of distinct inward and outward movement with spatial scales of $\sim$5-20 kpc. This is further visualized in Fig.~\ref{fig:ba_azimuthaldist}, which shows certain azimuthal angles experiencing strong radial inflows, while others have strong outflows or little net radial flow. Note, Fig.~\ref{fig:ba_azimuthaldist} was averaged over a relatively short period of time to emphasize the differences at a given redshift. Dips come from the radial oscillation of gas within and just outside the disk as well as anisotropies of accreting gas at the given time. On average, there is still a radial inflow signature for all galaxies.

Regions of similar size with higher infall or outflow rates are also seen in the vertical mass flux maps for galaxies with stronger vertical flow rates (\textbf{m12f}, \textbf{m12b}), while the galaxies with weaker flows tend to be less structured and more randomly distributed (\textbf{m12m}, \textbf{m12i}) (Figure \ref{fig:vertical_velocity_maps}). We do note, however, that we see isolated regions with high vertical velocities, as would be expected from areas with high SFRs that can drive galactic outflows/fountains (see Section~\ref{sec:discussion-observations}).

\subsection{Evolution In Disk}\label{sec:results-evolution-in-disk}

\begin{figure} 
	  \center{\includegraphics[width=.48\textwidth]
	       {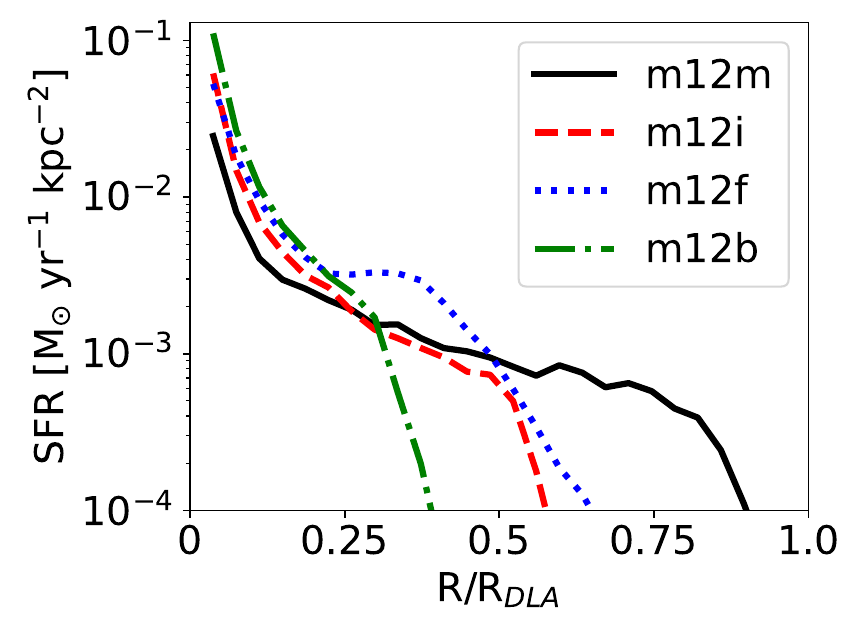}}
	  \caption{\label{fig:pt-sfh} Distribution of star formation rate per unit area as a function of cylindrical radius averaged for $0 \le z \le 0.2$. SFR per unit area peaks at the center for all four galaxies. The larger disk galaxies (\textbf{m12m}, \textbf{m12i}, \textbf{m12f}) show star formation at larger radii with a sharp drop off at the disk edge. Star formation in \textbf{m12b} is much more centrally peaked.
     }
\end{figure}

\begin{figure} 
	  \center{\includegraphics[width=.48\textwidth]
	       {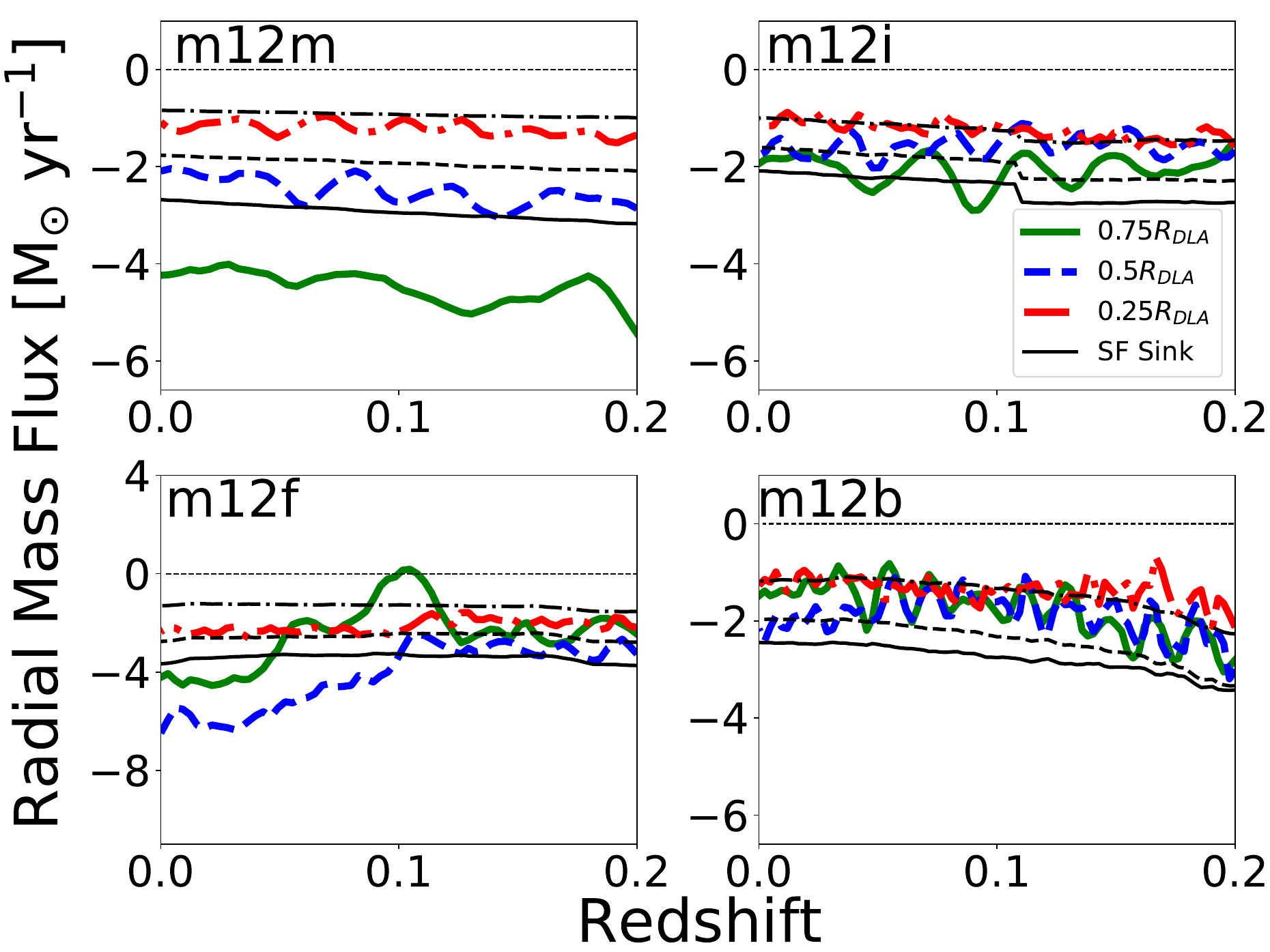}}
     \caption{\label{fig:ba-disk-mf}  Cylindrical radial mass flux through the disk plane. Mass flux was calculated in a 1 kpc radial bin at various radii within the disk, for gas within 2 scale heights of the disk plane. Star formation sink rates (black, thin lines) were calculated as the negative of star formation rate interior to 0.75 \rgalStop, 0.5 \rgalStop, and 0.25 \rgalStop. Both \rgal and the disk scale height were calculated at each time point as described in Table~\ref{table:disks} to account for changes in disk size. Curves were smoothed using a moving average with a filter width of 2 dynamical times in order to suppress oscillatory movement. The SFR curve was smoothed as well for consistency, but is relatively stable at late times \citep{muratov15}. Mass fluxes closer to the edge for most galaxies are around 2 -4 M$_{\odot}$/yr, which roughly corresponds to the star formation rate interior to this radius (Table~\ref{table:ba-mf}). In most galaxies, star formation rates are lower than mass flux through the disk, as some mass is lost in outflows.  Mass flux rates drop off as you go further into the galaxy as gas forms stars. Note that \textbf{m12f} undergoes a merger event at around redshift 0.1, resulting in heavy accretion and radial transport through the disk. 
     }
\end{figure}

Fig.~\ref{fig:pt-sfh} characterizes the radial distribution of star formation of each galaxy, showing that the bulk of star formation occurs at radii much smaller than $R_{\rm DLA}$. Given that most particles join the disk near the edge and most of the galactic star formation occurs deeply within the disk (Fig.~\ref{fig:pt-sfh}), there must be radial transport through the disk itself in order to sustain this star formation. Although the total star formation is peaked at the center of the galaxy, as particles transport radially in the disk they form stars at all radii, starting at \rgalStop. Fig.~\ref{fig:ba-disk-mf} shows a net radial mass flux rate of around 2 M$_{\odot}$/yr on average within the disk, and how mass flux through the disk drops off as a function of radius as gas is locked up in stars or ejected from the disk. This trend is not clear in \textbf{m12b} due to its very centrally concentrated SF. 

Our particle tracking analysis shows gas that joins in the outer regions of the disk can eventually transport to more interior star forming regions. The trajectories are not simple, however, with a radial oscillatory component related to spiral arm like structure. Averaging over the gas azimuthally renders this oscillatory motion to a slow overall radial inflow on the order of 1-3 km/s. It is worth noting, that when averaging individual particle trajectories over time, similar velocities are seen (Fig.~\ref{fig:pt-diskMovement}). This implies that the particles that actually fuel star formation in the interior regions can be accreted relatively long ago (i.e. it takes ~2 Gyr to travel 10 kpc at 5 km/s).

\section{Discussion}\label{sec:discussion}

\subsection{What determines the disk edge?}\label{sec:discussion-pileup}

\begin{figure} 
	   \center{\includegraphics[width=.45\textwidth]
	       {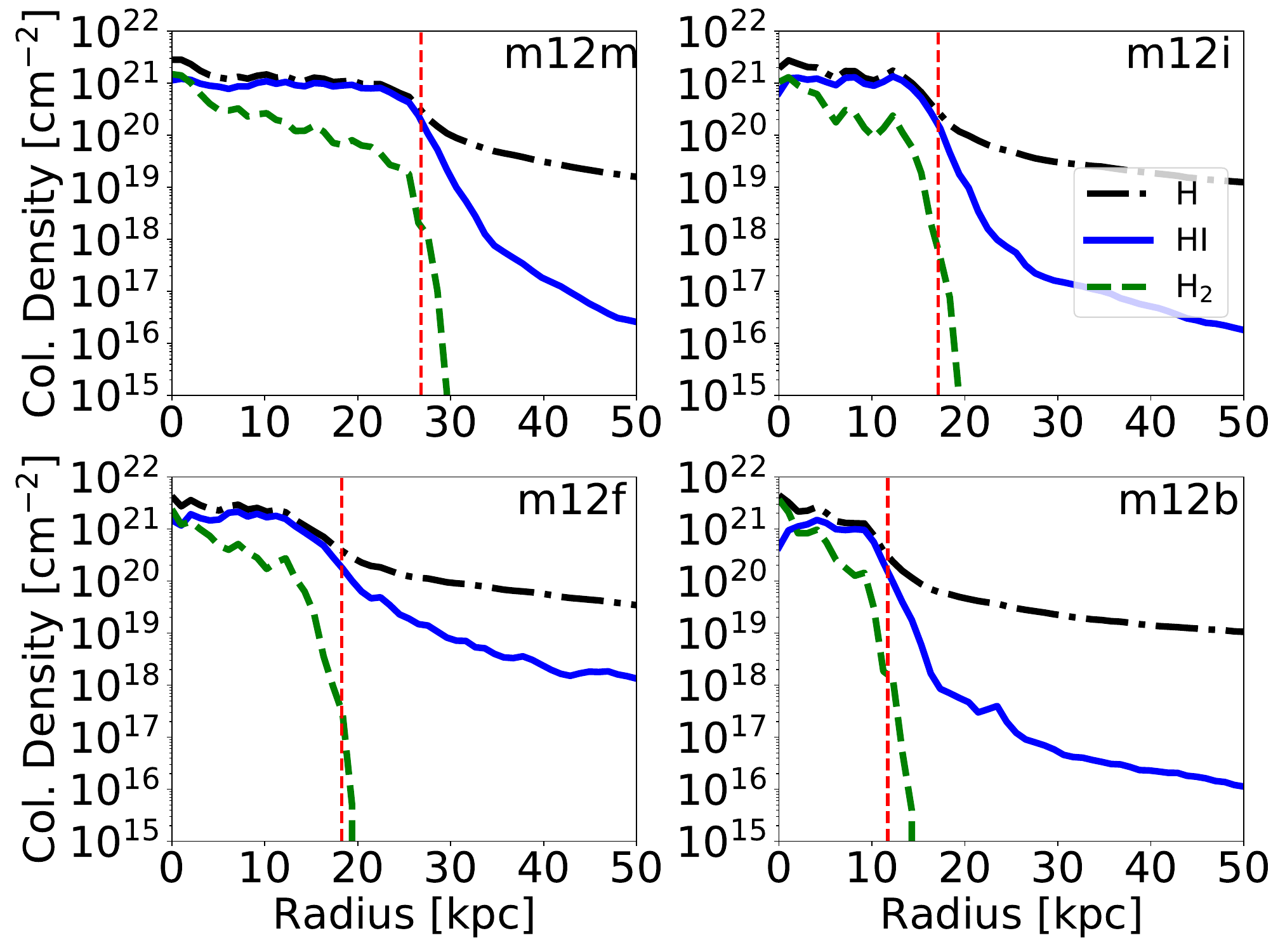}}
	  \caption{\label{fig:ba-ColDensQuant} Column density as a function of radius for total hydrogen, HI, and H$_{2}$ for the four galaxies in our sample viewed face on. Vertical dashed lines show \rgalStop. Note, galaxies show a sharp edge signified by a steep drop off in column density. This drop off is less extreme for total hydrogen, as a significant amount is ionized outside the disk.
	  } 
\end{figure}

Our simulations show gas "pile-up", i.e. accumulation, at the disk edge, marked by a decrease in average radial speed and ionization, and an increase in column density. 
In Fig.~\ref{fig:ba-ColDensQuant} we show that the change in column density and increase in neutral fraction occur at a similar radius as the change in radial speed (seen in Fig.~\ref{fig:ba-radvel_and_radmf}). This disk radius (\rgalStop) marks a physical "edge" of the gas distribution, with HI profiles similar to observed inclination-corrected profiles \citep{ianjamasimanana18}. Within  $\pm$20\% of this edge the total projected column density of gas changes by more than an order of magnitude, from less than $10^{20}\rm cm^{-2}$ to $\sim 10^{21} \rm cm^{-2}$. Because of this change, the gas also changes from mostly ionized to mostly neutral. We therefore conclude that the disk "edge" is caused by a physical accumulation of gas and the ionization change is simply a consequence of the strong density enhancement. Our findings in previous sections provide clear interpretation of this effect: low density, ionized gas accretes at relatively high radial velocities, while exhibiting an overall sense of co-rotation. As this gas joins the disk edge, it can no longer flow inwards as easily, causing it to slow down in radial direction, on average. This accumulation causes an increase in density and neutral fraction (owing to more efficient self-shielding). 

This interpretation of the disk edge marked in HI differs from many previous models, wherein the disk edge is primarily due to ionization change and the disk density structure is implicitly assumed to be much more extended \citep{corbelli93,schaye04,bland-hawthorn17,fumagalli17}. In these models, a break in the HI column density would be expected in the CGM at the photoionization limit ($\sim$ 5 $\times$ 10$^{19}$cm$^{-2}$). This break was not observed to be a common feature in galaxies in recent observations \citep{ianjamasimanana18}, which see a break at higher column densities, similar to our simulations, implying that ionization is not the main factor determining the extent of the HI disk. 

This steep drop in the column density of neutral hydrogen with increasing $r$ continues until it reaches column densities of $\sim 10^{17} \rm cm^{-2}$, typically within 1.5-2 \rgalStop. An exception is \textbf{m12f}, where a recent merger redistributes the gas, causing a more gradual drop that reaches $N_{\rm H}\sim 10^{18} \rm{cm}^{-2}$ within about 2 \rgalStop. These sharp drop offs suggest that disk outskirts at $z=0$ with column densities $\sim 10^{18}-10^{19}$ are not dramatically larger than the extent typically probed by current observations, i.e.$N_{\rm HI} >10^{19} \rm cm^{-2}$. This has consequences for detectability of lower density HI at $z\sim 0$ from very extended disks by ongoing and future observations (e.g., \citealt{pisano18,davis19,pingel19,witherspoon20}). Naturally, at higher redshifts, CGM gas has higher characteristic density, enabling more extended intermediate column density gas \citep{faucher16}. Finally, for rough comparison, we show the approximate column density of molecular hydrogen, which falls off even more sharply at the disk edge, with column densities falling from $10^{20} \rm{cm}^{-2}$ to less than $10^{15} \rm{cm}^{-2}$ within 30\% of \rgalStop. 

\subsection{Accretion from HVCs and IVCs}\label{sec:discussion-HVCs}

\begin{figure} 
	  \center{\includegraphics[width=.48\textwidth]
	       {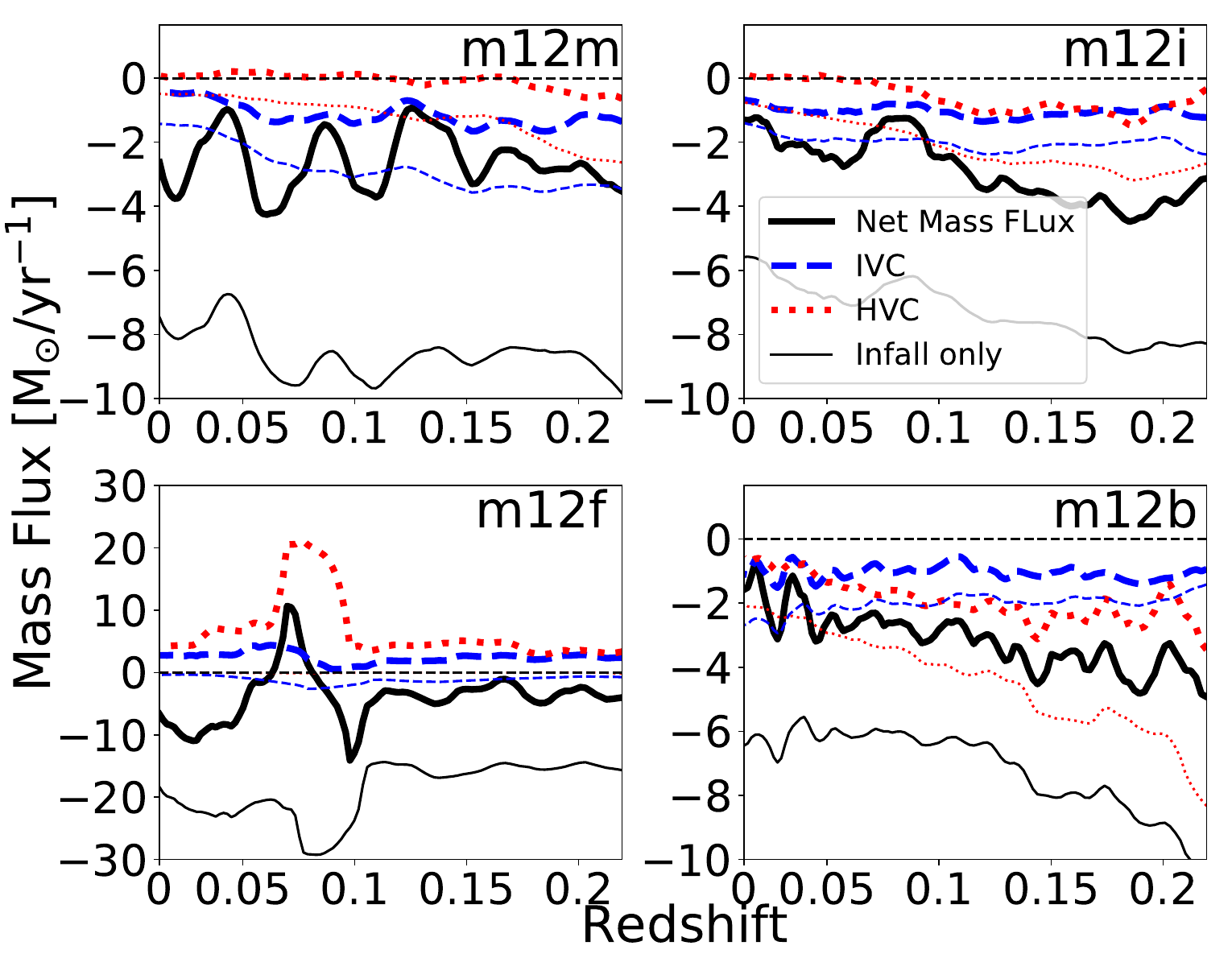}}
	  \caption{\label{fig:ba-hvc-mf} Gas mass flux (all gas, from HVCs only, and from IVCs only) through a cylindrical shell with a radius of \rgal and a height of $\pm$ 10 kpc. Negative values correspond to inflow. The thick lines shows the net mass flux from all gas, while the thin lines show only the mass flux from gas moving radially inward. HVCs and IVCs particles are identified by the deviation from galactic rotation curve as predicted from the enclosed mass as a function of  cylindrical radial position. The cutoff for HVCs is above 70 km/s. The cutoff for IVCs is from 40-70 km/s. This classification was done in order to approximate observations, however, it does not reflect what would actually be observed from within the galaxies. Instead, it serves as an estimate of the relative contributions of non-corotating gas.
}
\end{figure}

In order to roughly estimate accretion from gas within HVCs and IVCs kinematics in our simulations, we calculate their mass flux through a cylindrical surface with a radius of \rgal and a height of +/- 10 kpc (Fig.~\ref{fig:ba-hvc-mf}). We flag HVCs as any particle with deviations from the galactic rotation curve\footnote{We calculate the deviation of the circular velocity predicted by the enclosed mass interior to the particle's cylindrical radius on the disk and the tangential component of the particle velocity projected onto the disk plane.} above 70 km/s. IVCs are identified as particles with deviations of 40-70 km/s. In the non-merging large disks, \textbf{m12m} and {\bf m12i}, HVC mass fluxes range from 0-20\% ($\sim$ 0-1.5 \MFunit) of net accretion and 5-30\% ($\sim$ 0.5-3 \MFunit) of the gas infall rate.\footnote{Ratios were calculated simply as net mass flux from HVCs only divided by net mass flux of all gas.} The fraction of net mass flux in HVCs is decreasing with time and is negligible at $z \sim 0$.  IVC mass fluxes make up a similar percentage of accretion on average, but do not change as significantly with redshift. In \textbf{m12f}, a recent merger, there is little HVC or IVC accretion; instead this galaxy shows net HVC outflow. However in a more compact disk, \textbf{m12b}, HVC and IVC have similar contributions and jointly dominate the gas infall rate. Particles flagged as HVCs tend to have very low specific angular momenta (order $\vec{j}<\sim 1000$ kpc km/s) and rotate well below the rotation curve at their position.

We generally find that HVC mass flux makes up $0-30\%$ of the gas infall rate and of the net gas mass flux rate (net flux contribution is slightly higher in {\bf m12b}) at $z\sim 0$. This implies that most accreting gas is not in HVCs, even when total (ionized+ neutral) gas is taken into account, as we do in our analysis. This finding is in rough agreement with observational results of \cite{putman12} who found HVCs (total mass of all phases) account for roughly 13-20\% of accretion needed given the observed star formation rates. IVCs make up a similar, but slighly higher, fraction of the mass fluxes. \citet{Rohser18} found IVC accretion rates of 0.52 M$_{\odot}$/yr which is of a similar order as our results. The authors note, however, that they were limited to latitudes above 20 degrees and would expect more IVC accretion from an all sky survey. 

This analysis is only a crude estimate of the relative contribution of HVCs and IVCs to the galactic gas accretion at late times and is not perfectly analogous to what would be seen in observations. One important difference is that these values correspond to total gas mass flux, as opposed to neutral hydrogen mass flux (which can be an order of magnitude smaller). Additionally, we are simply counting mass flux through a cylindrical shell at the disk edge. If we track mass flux further out, the proportion of HVC flux increases as gas at larger radii begins to deviate from the rotation of the disk (see Fig.~\ref{fig:ba-j-graphs}), but at the same time, gas neutral fraction drops. Finally, when looking at accretion only (thin lines in Fig.~\ref{fig:ba-hvc-mf}), the relative proportions between HVCs, IVCs, and total accretion are similar, but values are higher as outflowing gas has similar kinematics. 

An additional caveat of our simple analysis is that we only measure global rates: local deviations that would be seen by an observer within the galaxy could be substantial. For example, different azimuthal angles in Fig.~\ref{fig:ba_azimuthaldist} show varying magnitudes of inflow and outflow that will also affect local population of IVCs and HVCs.

Similarly, one would need to know full 3D kinematic information to robustly determine the gas infall rates. While velocities of the approaching gas are broadly aligned with the disk plane, this alignment is not perfect. Fig.~\ref{fig:pt-trajectories} and Fig.~\ref{fig:pt-angles} show the distribution of polar angles at which gas approaches the disk. These particles typically approach at a small nonzero angle ($\sim$ 15$\degr$). There are few particles approaching the galaxies along the disk plane, and essentially no gas is approaching directly perpendicular to the disk. 
Interestingly, gas coming in at higher angles ($\theta > \sim 30^{\circ}$) tends to have lower specific angular momenta ($\sim$ 1000 kpc km/s lower) and tends to join the disk at smaller radii, avoiding the pile-up of gas at the disk edge (Sec.~\ref{sec:discussion-pileup}). These particles are more likely to correspond to HVCs and IVCs. 

Taking into account full comparison with observations from the perspective within the disk is beyond the scope of this paper and is planned for the future work. Furthermore we also plan to do a direct comparison with constraints from extra-planar gas kinematics in external galaxies that suggest more significant contribution from vertical gas infall \citep[e.g.][]{marasco19}.

\subsection{Radial Mass Flux and Star Formation in Disk}\label{sec:discussion-MF-and-SF}

 As discussed in Section~\ref{sec:discussion-pileup}, as gas joins the disk at large radii, the radial speed slows down to $\sim$ 1 km/s on average. The local motion of the gas in the disk is influenced by oscillatory effects from the spiral arms, as well as stellar feedback and to a lesser extent gravitationally powered turbulence \citep{orr19}. However, ring averaged net radial motion is inward, producing a net radial transport through the disk of 1-3 km/s and flow rates comparable to the disk star formation rates ($\sim 2-4$ \MFunit) as shown in Fig.~\ref{fig:ba-disk-mf} and Fig.~\ref{fig:pt-diskMovement}. 

The alternating velocity structure and related density enhancements show the presence of spiral arm like structure in the four galaxies in our sample. The structure in \textbf{m12m}, \textbf{m12i}, and \textbf{m12f} is floculent in nature, while \textbf{m12b} shows more distinct arms. While these oscillations complicate the radial transport of gas through the disk, it has been shown that spiral structure and density inhomogeneities can transport angular momentum to outer Linblad resonances \citep{bell79} on galactic scales. On smaller scales, \cite{quataert10,quataert11} showed that m=1 density inhomogeneities (single armed sprials) can drive gas inflow up to 10 \MFunit \hspace{.1cm} onto central AGNs.

The {\it net} accretion rate, SFR, and mass flux through the disk at large radii ($\sim 0.75$ \rgalStop) are similar when averaged over time (Table~\ref{table:ba-mf}), implying a balance between gas inflow and star formation rate. It is worth noting that while the above is true for the net accretion, the total gas infall rates are significantly higher (see Fig.~\ref{fig:ba-hvc-mf}) because of outflowing gas component caused by the oscillatory motion and CR driven outflows from galaxies \citep{hopkins21_crOutflows, chan_inPrep}. \cite{goldbaum15,goldbaum16} similarly studied radial mass fluxes in simulations of galactic disks with and without stellar feedback. In their fiducial runs with stellar feedback, they found mass fluxes on the order of 0.5-1 \MFunit with SFRs of 2-3 \MFunit. Unlike this study, these simulations were non-cosmological and looked at disks in earlier stages, before equilibrium is reached. When equilibrium is reached, the SFR may more closely match the radial mass flux at all radii, as we see in our study.

As gas transports inwards it becomes locked up in stars, resulting in the overall mass flux rate dropping with decreasing radius. This can be seen clearly in Fig.~\ref{fig:ba-radvel_and_radmf} and \ref{fig:ba-disk-mf}. In general, infall rate can be slightly higher than star formation rate owing to galactic outflows that can eject some of the gas supplied to the disk. Depending on the distribution of star formation in each galaxy (Fig.~\ref{fig:pt-sfh}), we expect to see differences in how this drop off behaves as a function of radius. While in all galaxies the star formation per unit area peaks at small radii, the total star formation in \textbf{m12m} is more evenly distributed throughout the disk. Consequently, in Fig.~\ref{fig:ba-disk-mf} we see roughly constant drop in mass flux between radii, reaching a much smaller value at the smallest radius. Alternatively, \textbf{m12i} shows a stronger peak of star formation in central region, resulting in a smaller change between the inner and outer radii on average. In a more extreme case, \textbf{m12b} has almost all star formation in the center-most region due to its small disk size. Correspondingly, the mass fluxes through the entire disk are roughly equivalent, only decreasing slightly on average.

 While average radial infall speeds seen in azimuthal rings or time averaged gas motion are low (see Fig.~\ref{fig:ba-radvel_and_radmf},~\ref{fig:ba-disk-mf}), the process of radial infall still requires slow, but continuous net angular momentum loss. We will explore the physics of this process in future work. Additionally, it is worth noting that our average inflow velocities are broadly consistent with what is predicted by theoretical models of ``gravito-turbulent'' disks or disks with an effective ``turbulent viscosity'' \citep{gammie01,rice05,cossins09,hopkins13}, which (for supersonic, isotropic turbulence) predict inflow rates $\dot{M} \sim 2\pi\,\sigma_{\rm turb}^{2}\,\Sigma_{\rm gas}/\Omega$ \citep{thompson05}, or net radial inflow velocities $\langle v_{r} \rangle \sim (\sigma_{\rm turb}/V_{c})\,\sigma_{\rm turb}$. For a velocity dispersion here of $\sigma_{\rm turb} \sim 10-25\,{\rm km\,s^{-1}}$ \citep{chan_inPrep} and circular velocity $V_{c} \sim 200\,{\rm km\,s^{-1}}$ predicts $\langle v_{r} \rangle \sim 0.5-3\,{\rm km\,s^{-1}}$. This suggests a steady, ``gravito-turbulent'' effective angular momentum loss rate on large scales may be a reasonable approximation.

\subsection{Additional Observational Implications}\label{sec:discussion-observations}

In this section we comment on the viability of observing accretion near the disk edge in cases of more stable disks (\textbf{m12m} and \textbf{m12i}) and disks with heavier accretion (\textbf{m12f} and \textbf{m12b}), as well as radial gas flows within the disk. HI column densities remain above $10^{19} \rm cm^{-2}$ during the transition between the relatively uniform, moderate velocity flow close to the disk edge and much lower average radial velocities inside the disk (Fig.~\ref{fig:ba-ColDensQuant}), making such a transition readily observable in HI emission as well as absorption in HI and associated metal species.

Outside the disk in more stable galaxies (\textbf{m12m} and \textbf{m12i}), we expect to see inflow velocities peaking around 30-40 km/s with infalling region length scales of 1-10 kpc (Fig.~\ref{fig:ba_azimuthaldist}). We predict a ring averaged inflow velocity for this region of around 10 km/s. \citet{werk19} and \citet{bish19} have mapped absorption signatures toward a sample of seven blue horizontal branch stars above the Milky Way's disk using data from the Cosmic Origins Spectrograph (COS; \citealt{froning09}) and Keck/HIRES \citep{vogt94}. They found coherent ionized structures at the disk-halo interface with scale lengths of at least 1 kpc and radial velocities of 30-50 km/s, which is comprable to our expectations. Given the survey geometry (looking for absorption above the disk not far from the Solar circle) it is likely that the dominant gas accretion component would be missed by such a survey. As shown in Fig.~\ref{fig:pt-trajectories} and Fig.~\ref{fig:pt-angles}, the average angle of approach is small. We note, however, that gas approaching at higher angles tends to increase the angle at which it joins the disk (i.e. falls onto the disk) just prior to joining. The local angle of accretion for this gas would therefore be higher than these figures may imply. In future work we plan to do a more detailed comparison with from the Solar circle  perspective in our simulated galaxies.

 \cite{ho20}  used Mg II absorbers from background quasars near z=0.2 galaxies \citep{ho17,martin19} coupled with galaxy orientation based on high resolution images to constrain maximum radial inflow speeds to 30-40 km s$^{-1}$ along two lines of sight in separate galaxies (at galactocentric radii of 69 and 115 kpc). However, along other lines of sight, they are not able to successfully model radial inflow in agreement with their constraints on disk orientation, leading the authors to conclude there is no radial inflow. While their sight-lines probe further from the galaxies than the region of focus in this work, these results are in agreement with both our numerical values for inflow velocities, as well as anisotropic accretion around the simulated disk galaxies, with some regions having no inflow at a given time.

For galaxies undergoing more active accretion (\textbf{m12b} and \textbf{m12f}), we predict a peak inflow rate as high as 50 and 100 km/s respectively, with ring averaged velocities around 15 km/s. \citet{zheng17} observed 7 UV bright disk stars in M33 as background sources using the COS in order to identify significant ionized gas accretion. Accretion velocity was $v_{\rm acc}=100^{+15}_{-20} \rm{ km s^{-1}}$ and estimated mass flux was 2.9 M$_{\odot} \rm{yr}^{-1}$ (5.8 M$_{\odot} \rm{yr}^{-1}$ if flows are axisymmetric). While these are within the range of values for our simulated MW-mass galaxies, M33 has significantly lower mass and is therefore expected to have lower (long-term averaged) infall rate. The authors note that this high mass flux rate is likely transient, either due to reaccretion from a large galactic fountain event, or recent interaction with M31. Note that total infall rates are higher than net accretion rates owing to infall/outflow variations in and around the disk. Extrapolation of individual regions into disk averaged quantities is therefore very uncertain.

Within the disk, local gas motion (infall and outflowing gas) can be strongly affected by the oscillations caused by the spiral arms, as well as local feedback and collapse, often driving the gas up to 30-40 km/s. 
However, when ring averaged, our simulations predict radial gas inflow on the order of 1-3 km/s (see Fig.~\ref{fig:ba-radvel_and_radmf} and Fig.~\ref{fig:pt-diskMovement}). The difficulty of determining radial velocities from a single orientation complicates observation of these flows. If galaxies are viewed perfectly face-on, radial velocities would be impossible to detect. If viewed perfectly edge on, it would be difficult to spatially resolve different regions in the disk so additional modeling of the observed gas velocities is needed.

\cite{wong04} searched for radial flow signatures within the disk using CO and HI velocity fields in seven nearby spirals by decomposing the velocity fields of concentric elliptical rings into a third-order Fourier series. For three galaxies, an upper limit of 5-10 km s$^{-1}$ for inflow velocities were determined. These limits are consistent with our ring averaged inflow velocities (Fig.~\ref{fig:ba-radvel_and_radmf}). For our simulated galaxies ring averaged velocities within 0-0.5\rgal are very small, ~1-3 km/s depending on the ring and galaxy, while values can be larger in the outskirts.\footnote{Two galaxies in the \cite{wong04} sample also showed signs of radial outflows at 45 and 60 km s$^{-1}$ (with similar uncertainties as the inflow signatures), which match the instantaneous radial velocities seen in Fig.~\ref{fig:radvel-maps}. For some galaxies, inclinations were too low or uncertain to be able to identify any radial flows.}

A similar study using 10 galaxies from the THINGS survey with stellar masses comparable to our sample \citep{schmidt16} found radial inflows signatures as a function of radius for a few of these galaxies. Inside the optical radius for certain galaxies, the authors found ring-averaged radial inflow velocities on the order of $\sim$ 5-20 km s$^{-1}$ with corresponding mass fluxes of $\sim$ 0.1-4.0 \MFunit depending on the radius within each galaxy. This uncertainty range is in general agreement with out findings when time variations in accretion rate and velocity are accounted for. Contrary to what was expected from their simplified model of galaxy formation, the mass fluxes varied strongly as a function of radius, instead of steadily decreasing with radius based on the interior star formation rate. This is likely due to the fact that they are only able to observe a single time point, which will be confounded by spiral structure oscillations and outflow events. As shown in the bottom row of Fig.~\ref{fig:ba-radvel_and_radmf}, dispersion over time and radial variations in the mass flux are much larger than average values. 

\subsection{Comparison to Simulations without Cosmic Rays}
\label{sec:CR-hydro}
In this section we briefly summarize the differences between the simulations used in this study \citep{hopkins20} (CR+), simulations with the same model but omitting CRs, (MHD+), and previous FIRE-2 simulations without MHD and cosmic ray physics (Hydro+) \citep{hopkins18}. Further analysis can be found in Appendix~\ref{sec:appendix_comparisons}.

The addition of cosmic rays has an effect on the specific quantitative results of our paper, however, the qualitative picture remains roughly the same. The higher infall rates in the Hydro+ and MHD+ runs (Fig.~\ref{fig:radMassFluxQuantComp}) drive a factor of 2-3 higher star formation within the disk \ref{table:mf-comp}. Feedback energy and momentum input, as well as gravitationally driven turbulence cause higher disk velocities. Velocity dispersions within the CR+ and Hydro+ runs are 10-25 km/s and 15-30 km/s respectively \citep{orr19,gurvich20,chan_inPrep}. This can also be seen in the larger flow velocities within the disk in the Hydro+ runs, both in the extrema at a given time point ($\sim$ 60 km s$^{-1}$) (Fig.~\ref{fig:comp_RadVelMaps}) and on average ($\sim$5-10 km s$^{-1}$ averaged azimuthally) (Fig.~\ref{fig:radVelQuantComp}).

The interaction between the CRs and gas outside the disk is the primary cause of the above differences. Gas outside the disk in the CR runs tends to be falling in at lower velocities. Additionally, \cite{hopkins21_crOutflows} shows that CR driven winds are preferentially driven perpendicular to the disk plane, confining infalling gas to closer alignment with the plane. By comparison, the Hydro+ runs tend to have higher infalling velocities in the inner CGM (Fig.~\ref{fig:radVelQuantComp}), and approach the disk plane from larger angles on average (Fig.~\ref{fig:comp_jz}). The dominant inflow is still parallel to the disk on average, however the mass fluxes in the Hydro+ and MHD+ runs vary more strongly over time (Fig.~\ref{fig:comp_mf_directions}). The nature of gas accretion at late times as well as its thermal properties and connection to the cooling flow solution in Hydro+ simulations will be further explored in \cite{hafen_inPrep}.

 While there are observational constraints (specifically gamma ray emission) for CR transport on galactic scales that were used to select a default value for the constant diffusion coefficient, $\kappa_{||} = 3\times 10^{29} \rm cm^2/s$,  \citep{chan19, hopkins20}, there are currently no direct constraints on the CRs in the CGM. The constant diffusion model in our simulations is the simplest possibility, however there are more physically motivated models for CR diffusion and streaming that we have recently tested in \cite{hopkins_21_crCoeff}. When constrained to match observations on galactic scales, these models result in higher effective diffusion within the CGM, suggesting lower overall CR energy density and pressure gradients. The outcome is that the resulting effects on the CGM, SFRs and galactic outflows in these models lie in-between our Hydro+ runs and our CR runs with constant diffusion \citep{hopkins_21_crTransport}. We expect similar conclusions to apply for the gas infall. In this respect Hydro+/MHD+ and CR+ runs represent two different limiting cases describing the nature of gas infall on galactic scales.

\section{Conclusions}

We analyse gas accretion in and around four low redshift $L_*$ disk galaxies simulated using the FIRE-2 model with feedback from cosmic rays. Galaxies span a wide range of disk sizes. We find the dominant source of accretion is gas that is largely co-rotating and joining near the gaseous disk edge. Galactic  gas then radially flows inward towards regions with active star formation with low (time and azimuthally averaged) velocities of few km/s.

The key points of our paper are summarized below:

\begin{itemize}
  \item Net accretion rate onto the galaxy roughly corresponds to radial gas mass flux through the disk and total star formation rate of late time MW-mass galaxies.
  
  \item Infalling gas within tens of kpc from the disk is moving primarily parallel to the disk. Infalling mass flux in the direction vertical to the disk is subdominant.
  
  \item Gas in the inner CGM is largely co-rotating with the disk with specific angular momentum similar to the disk edge but below value needed for full rotational support. As the gas moves radially and accretes to the disk, it largely conserves angular momentum with only a small loss until it reaches full rotational support, typically just after crossing the gaseous disk edge.
  
  \item Because infalling gas largely corotates, the ring averaged infall velocity 30-40 kpc from the center of the non-interacting MW-mass disks is only $\sim$ 20 km/s, a small fraction of rotational velocity. Time averaged radial velocities drop to $\sim 10$ km/s for (20-30km/s for a disk with recent interaction) at $\sim \rm R_{\rm DLA}$, radius at which neutral gas column density reaches DLA limit, which we take to be the disk edge. This effect and strong alignment of the gas infall with the disk, make it difficult to observationally  quantify gas infall in the inner CGM. Our results suggest future observations of gas around local MW-mass star forming disks galaxies to include corotating gas near the disk when accounting for the galactic gas accretion.
   
 \item Because the dominant source of accretion is gas that is largely co-rotating, the gas with HVCs and IVCs velocities is typically a sub-dominant component.

  \item Accreting gas tends to "pile up" at the disk edge. As gas from the CGM joins the disk, its radial speed drops from 10-20 km/s to 1-3 km/s and density increases, causing a fast change in its ionization state. This has consequences for the observability of the extended neutral gas structures well outside \rgalStop. 

  \item Gas accretes at small nonzero angles, typically 5-20$^\circ$ above or below the disk when viewed one orbital period before joining the disk. Gas coming in at larger angles tend to fall onto the disk edge prior to joining, showing larger vertical velocities during the process.
  
  \item Most gas joins the disk close but slightly interior to the gaseous disk edge. In all cases this is well outside the bulk of the stellar disk, typically at $\sim 3R_{\rm *, 1/2}$.

  \item Once in the disk, radial movement of the gas is complex.  In addition to local disturbance by feedback and collapse, there is larger scale motion likely due to spiral arm like structure. Despite relatively high instantaneous radial velocities (up to $\sim$20-40 km/s), time and azimuthally averaged bulk  radial transport has net velocities of only several km/s.
  
  \item Observations of radial flows within the disk will need to be handled with care, keeping in mind the time and spatial variations that dominate over long term average trends. Clear signal of the radial gas flow requires $\sim \rm{km/s}$ accuracy from large sample of disk galaxies. 
  
  \item Gas infall in simulations without cosmic rays is qualitatively similar. However, the infall velocities and gas infall rates are on average higher than in simulations with cosmic rays (up to 50 km/s at distance twice the disk edge and infall rates of $\sim$5 \MFunit), with much higher time variations in both.
  
\end{itemize}

 In future work we plan to make more direct comparison of our simulations with observations of the Milky Way and resolved disk galaxies at $z\sim0$. This will include mock observations of simulated HI emission and absorption spectra of different ions for spatially resolved "down the barrel" analysis, mock observations from the "solar circle" perspective, and the analysis of extra-planar gas kinematics.

 We will also investigate angular momentum evolution of gas as it accretes and flows within galaxies, including the physical origin of the related angular momentum loss. 

We note that most of our findings are qualitatively similar in simulations with and without CRs but quantitative differences in gas velocities and accretion rates can be significant. Difference in gas phase structure are also significant \citep{ji20}. We therefore plan to analyze simulations with a broader set of CR transport models that span the range between models with constant diffusion of CRs and Hydro+ (see section  \ref{sec:CR-hydro}).

\section{Acknowledgements}
DK was supported by NSF grants AST-1715101 and AST-2108314, and the  Cottrell Scholar Award from the Research Corporation for Science Advancement. CAFG was supported by NSF through grants AST-1715216 and CAREER award AST-1652522; by NASA through grant 17-ATP17-0067; and by a Cottrell Scholar Award from the Research Corporation for Science Advancement. Support for PFH was provided by NSF Research Grants 1911233 \&\ 20009234, NSF CAREER grant 1455342, NASA grants 80NSSC18K0562, HST-AR-15800.001-A. Numerical calculations were run on the Caltech compute cluster ``Wheeler,'' allocations FTA-Hopkins/AST20016 supported by the NSF and TACC, and NASA HEC SMD-16-7592. TKC was supported by the Science and Technology Facilities Council (STFC) through Consolidated Grants ST/P000541/1 and ST/T000244/1 for Astronomy at Durham. IE was supported by a Carnegie-Princeton Fellowship through the Carnegie Observatories. AW received support from NSF CAREER grant 2045928; NASA ATP grants 80NSSC18K1097 and 80NSSC20K0513; HST grants GO-14734, AR-15057, AR-15809, and GO-15902 from STScI; a Scialog Award from the Heising-Simons Foundation; and a Hellman Fellowship. The simulations presented here  used  computational  resources granted  by  the  Extreme  Science  and  Engineering  Discovery  Environment  (XSEDE),  which  is  supported  by  National  Science Foundation  grant  no.  OCI-1053575,  specifically  allocation  TG-AST120025 and resources provided by PRAC NSF.1713353 supported by the NSF. This work also made use of MATPLOTLIB \citep{matplotlib}, NUMPY \citep{numpy}, SCIPY \citep{scipy}, and NASA's Astrophysics Data System. We would like to thank the Kavli Institute for Theoretical Physics, supported in part by the National Science Foundation under grant No. NSF PHY-1748958,
and the participants of the {\it Fundamentals of Gaseous Halos} program for interactions that improved this work.

\section{Data Availability}
The data supporting the plots within this article are available on reasonable request to the corresponding author. A public version of the GIZMO code is available at http://www.tapir.caltech.edu/~phopkins/Site/GIZMO.html.

Additional data including simulation snapshots, initial conditions, and derived data products are available at
http://fire.northwestern.edu.

\bibliography{mybib}


\appendix

\section{Comparisons Between CR+, Hydro+, and MHD+ Runs}\label{sec:appendix_comparisons}

 As discussed in Section~\ref{sec:simulations}, the FIRE-2 simulations with MHD and cosmic ray physics (CR+) were used in this study due to their more realistic star formation rates and radial velocities compared with the hydrodynamical only simulations (Hydro+) and the MHD runs without the additional CR physics (MHD+) \citep{hopkins20}. In this appendix we highlight some key differences between the Hydro+, MHD+, and CR+ runs to illustrate that the overall qualitative properties of the gas flows near and inside galactic disks are the same. The three simulation suites are run with the same baryonic mass resolution of $m_{\rm gas}=7100 M_{\odot}$, and the same maximum spatial resolution and gravitational softening.

\begin{figure*}
	   \center{\includegraphics[width= 0.9 \textwidth]
	       {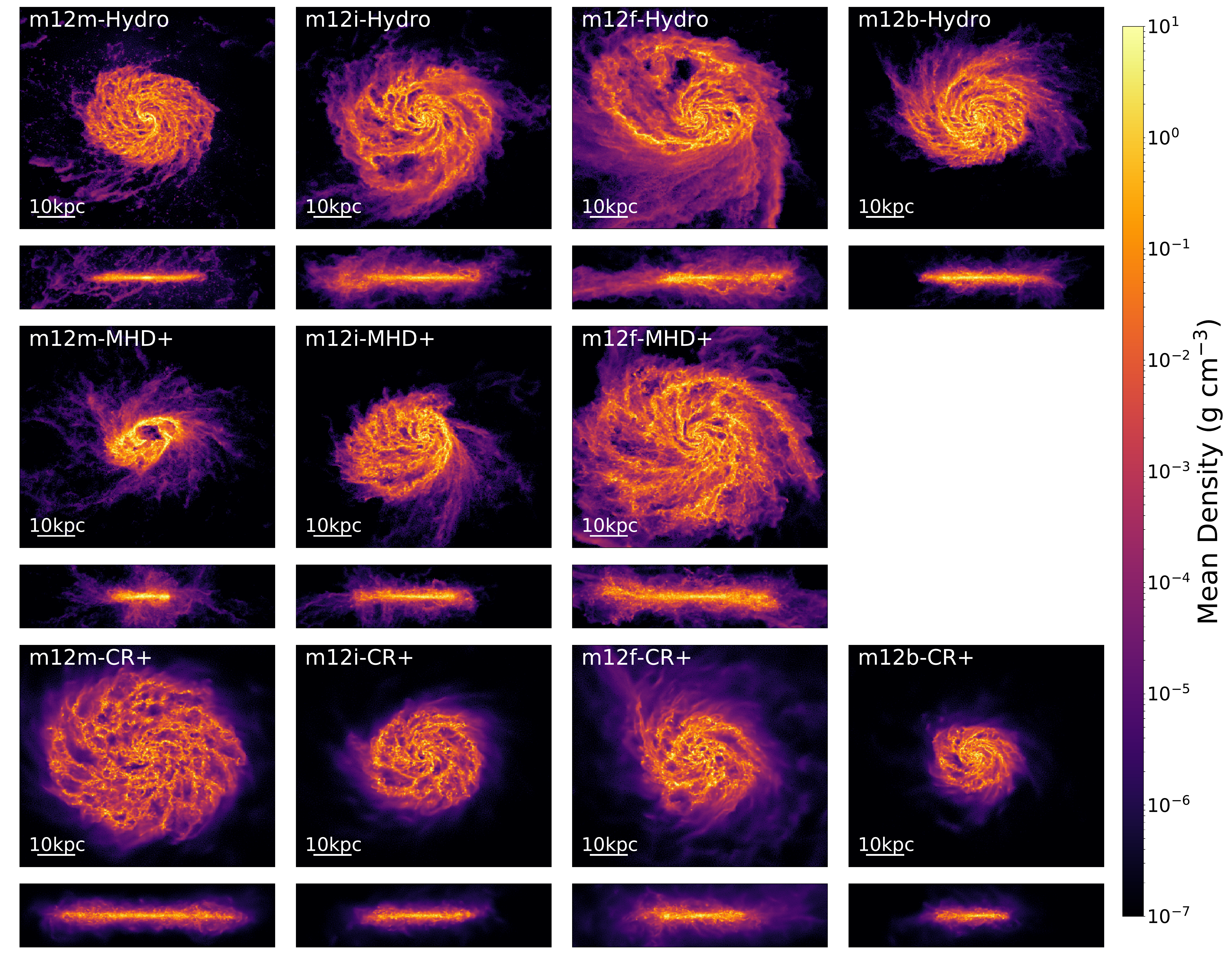}}
	 \vspace{.1 in}
	  \caption{\label{fig:comp_dens} Face-on view of average gas density at z=0. Hydro+ runs are shown on top, runs with MHD but not CRs (MHD+) are shown in the middle, and CR+ are shown on the bottom. The CR+ runs tend to form more cohesive disks on average.}
\end{figure*}

\begin{figure*}
	   \center{\includegraphics[width= 0.9 \textwidth]
	       {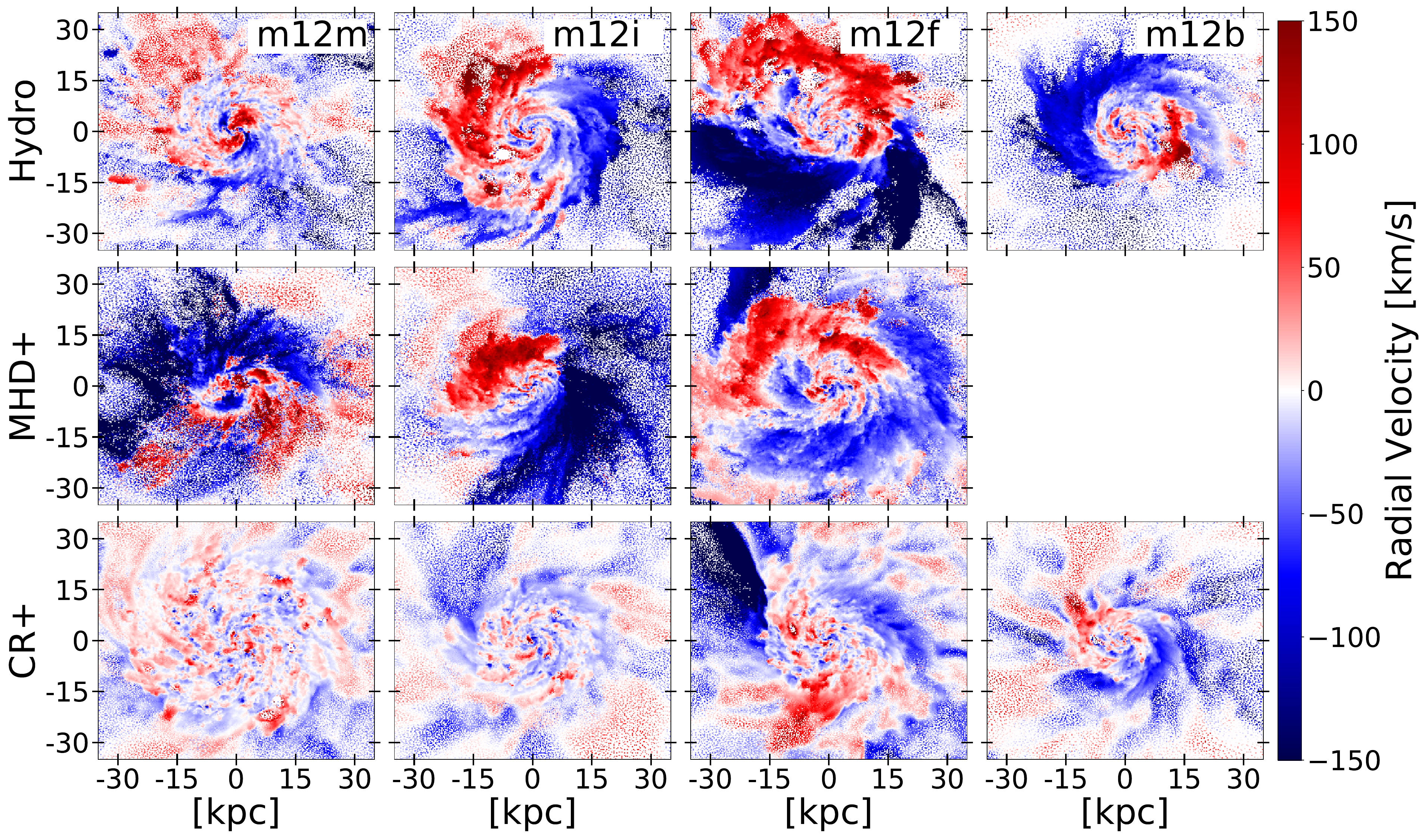}}
	  \vspace{.1 in}
	  \caption{\label{fig:comp_RadVelMaps} Radial velocity maps for the Hydro only runs (Hydro+, top), runs with MHD but not CRs (MHD+, middle), and the CR+ runs (top). Radial velocities in the Hydro+ runs tend to be 2-3 times higher on average, leading to stronger gas flows and less coherent disks.}
\end{figure*}

All sets of simulations form flat disks with spiral arms, however there are differences in gaseous and kinematic structure (Fig.~\ref{fig:comp_dens}). In general, CR pressure in CR+ runs smooths extreme over- and under-densities outside galaxies. The disks in the CR+ runs also are kinematically colder, owing to lower star formation rates and infall rates in the disk. Differences are clearly seen for \textbf{m12i} and \textbf{m12f} runs with and without CRs. The Hydro+ and MHD+ runs in these cases have large gaps in the gas distribution, arising from the extreme outward flowing radial velocities that can be seen in Fig.~\ref{fig:comp_RadVelMaps}.

\begin{figure*}
	   \center{\includegraphics[width= 0.9 \textwidth]
	       {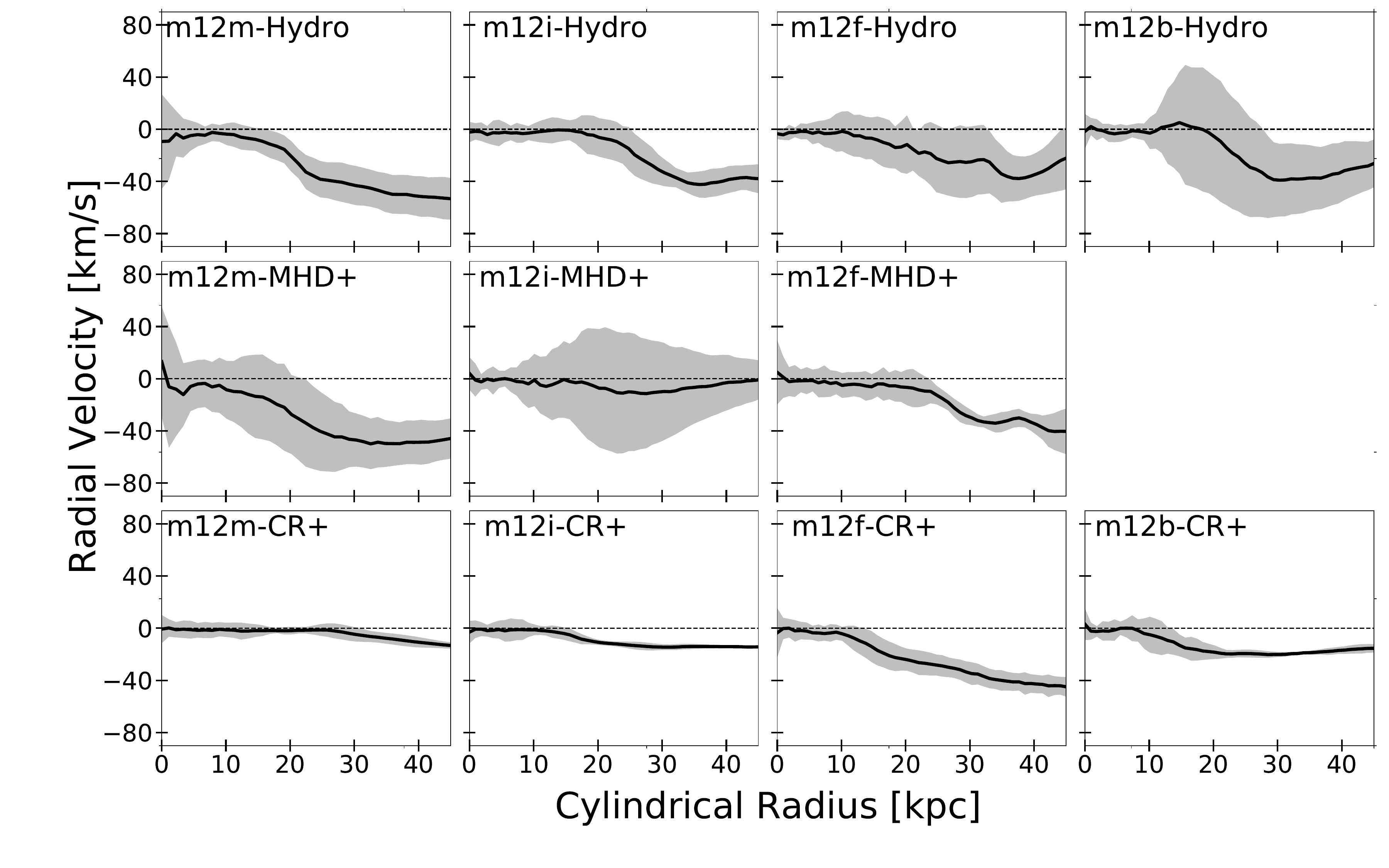}}
	  \vspace{.1 in}
	  \caption{\label{fig:radVelQuantComp} Cylindrical radial velocity as a function of cylindrical radial distance from the disk center for runs with Hydro only (Hydro+, top), with MHD but not CRs (MHD+, middle), and the CR+ runs (bottom). Values were averaged between $\pm$10 kpc as was done in Fig.~\ref{fig:ba-radvel_and_radmf}. Runs without cosmic rays tend to have more extreme radial velocity values on average, as well as larger deviations. The same general trend of gas slowing down as it reaches the disk edge remains between both runs.
}
\end{figure*}

\begin{figure*}
	   \center{\includegraphics[width= 0.9 \textwidth]
	       {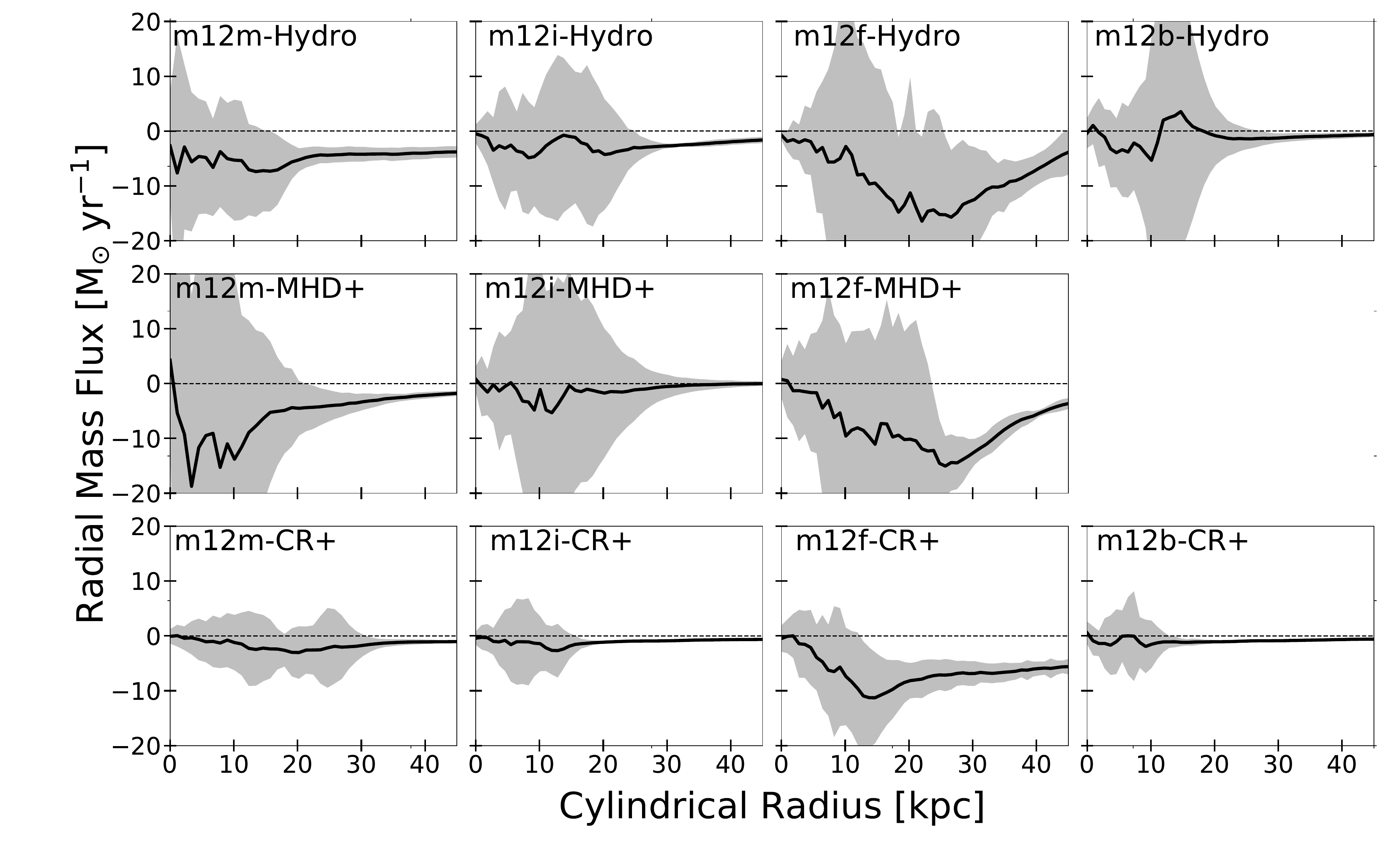}}
	  \vspace{.1 in}
	  \caption{\label{fig:radMassFluxQuantComp} Cylindrical radial mass flux as a function of cylindrical radial distance from the disk center for runs with Hydro only (Hydro+, top), with MHD but not CRs (MHD+, middle), and the CR+ runs (bottom). Values were averaged between $\pm$10 kpc as was done in Fig.~\ref{fig:ba-radvel_and_radmf}. Runs without cosmic rays tend to have more extreme radial mass fluxes on average, as well as larger deviations.
}
\end{figure*}

Similar to the CR+ runs, the Hydro+ and MHD+ runs show gas accreting parrallel to the disk and piling up at the disk edge (Sec.~\ref{sec:discussion-pileup}). The Hydro+ and MHD+ runs, however, show significantly higher accretion rates, mass fluxes, and star formation rates (Table~\ref{table:mf-comp}). Additionally, the drop in radial speed at the disk edge, which signifies this pile-up, is several times higher in the Hydro+ and MHD+ runs (Fig.~\ref{fig:radVelQuantComp},\ref{fig:radMassFluxQuantComp}). This is also obvious in Fig.~\ref{fig:comp_RadVelMaps}, where radial velocities within the disk and outside the disk are more extreme. The differences in these velocities ultimately arises from the additional pressure support provided by the cosmic rays, which becomes important in the CGM \citep{chan19,ji20, hopkins20}. This regulates gas flows in galactic halos with the net effect of slowing down the rapid gas infall. An indirect effect of lower infall rates are lower gas surface densities and star formation rates, resulting in more quiet kinematics of galactic disks (Chan et al., in preparation). 

\begin{figure*}
	   \center{\includegraphics[width= .9 \textwidth]
	       {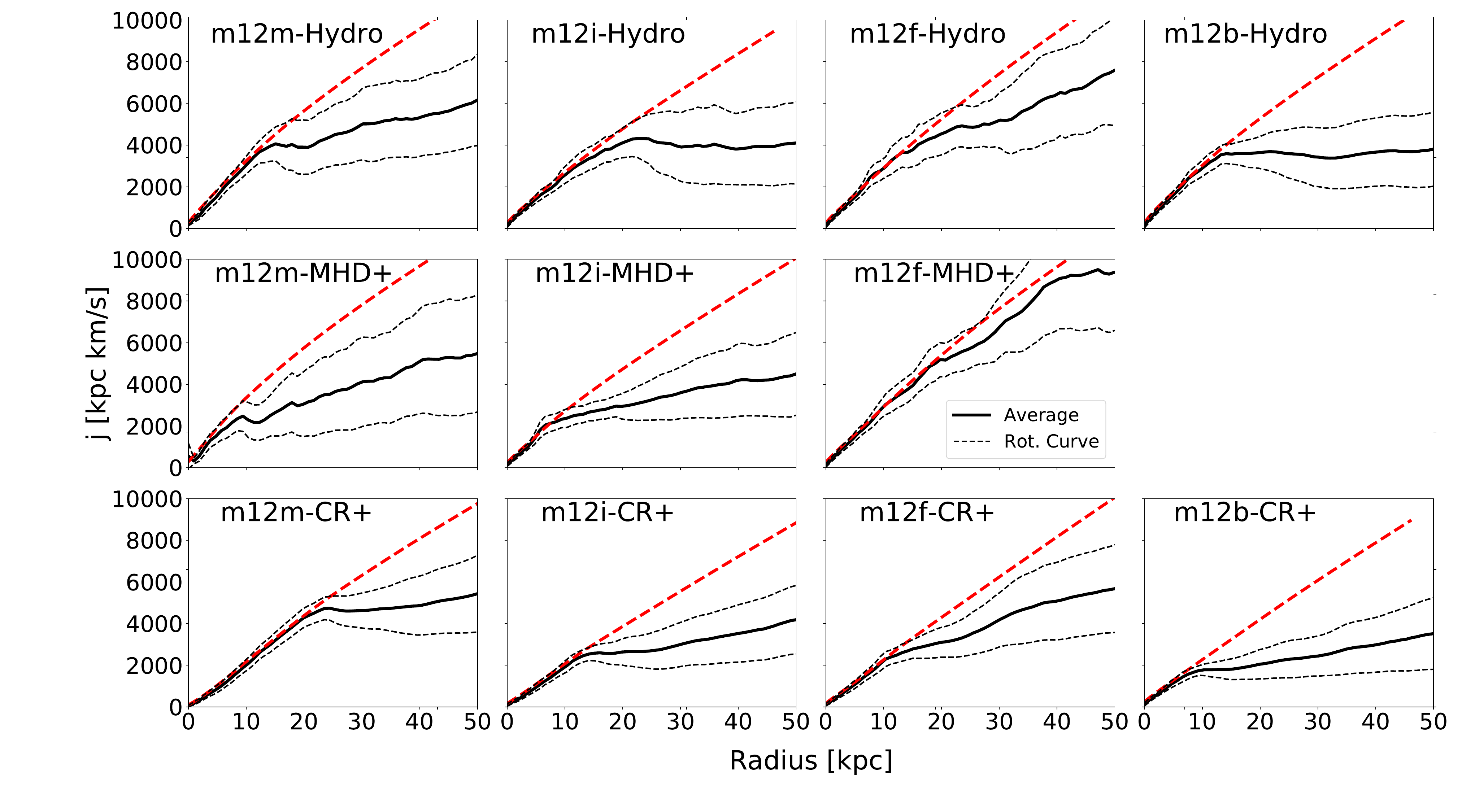}}
	  \vspace{.1 in}
	  \caption{\label{fig:comp_j_total} Specific angular momentum curves dependence on galactocentric radius for runs with Hydro only (Hydro+, top), with MHD but not CRs (MHD+, middle), and the CR+ runs (bottom). Thick red dashed lines show the specific angular momentum needed for rotational support. Thin black dashed lines show the standard deviation around the mean. All simulation suites show a tight distribution at r$\leq$\rgal where gas is fully rotationally supported. At larger radii, gas specific angular momentum is lower than needed for rotational support (as expected for the gas that is infalling) and deviations amongst particles at fixed radius are larger.}
\end{figure*}

\begin{figure*}
	   \center{\includegraphics[width= .9 \textwidth]
	       {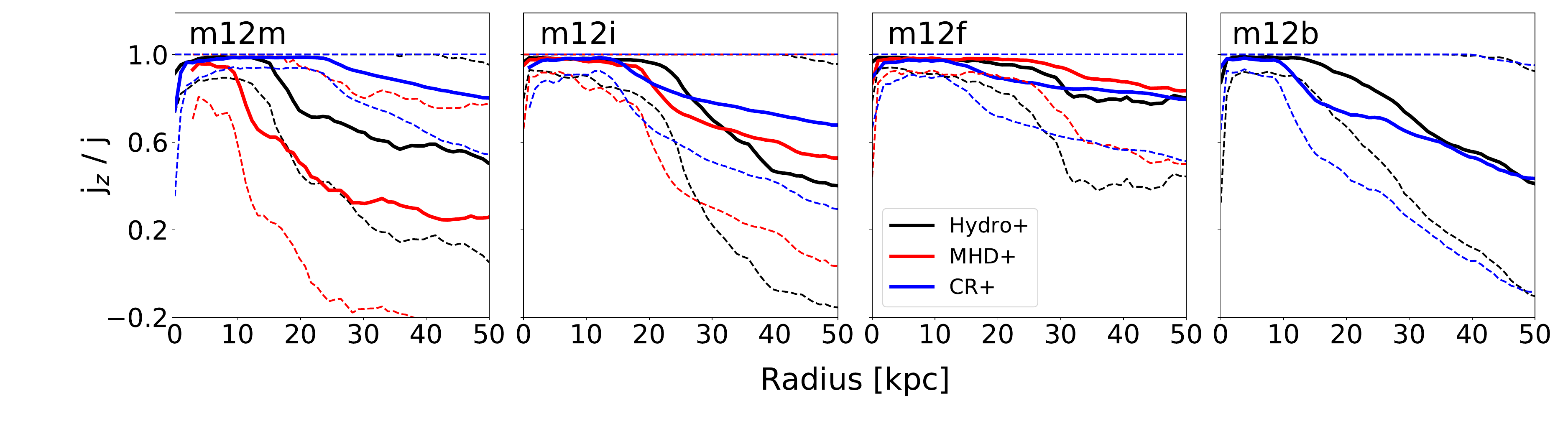}}
	  \vspace{.1 in}
	  \caption{\label{fig:comp_jz} Normalized z-components of the specific angular momenta: fraction of the total angular momentum in each radial bin in the direction of galactic angular momentum for runs with Hydro only (Hydro+, black), with MHD but not CRs (MHD+, red), and the CR+ runs (blue). Dashed lines show the standard deviation around the mean. Runs with cosmic rays tend to corotate more strongly on average, even out to large radii.
	  }
\end{figure*}

The structure of accreting gas is qualitatively the same in both runs, with gas accreting onto galactic disk along a funnel shape at relatively small angles with respect to the disk. Gas on large scales tends to be more strongly aligned with the plane of the disk in CR+ runs \citep{hopkins21_crOutflows}. The specific angular momentum of the gas particles in the Hydro+ and MHD+ runs follow closely the rotation curve for r$<$\rgalStop, while for r$>$\rgal gas starts rotating slower than is needed for angular momentum support (Fig.~\ref{fig:comp_j_total}). More significant differences are seen in Fig.~\ref{fig:comp_jz}. While the orientation of particle rotation is strongly co-aligned with galactic rotation within the disk and still loosely aligned at higher radii, it drops off more sharply and has much larger spread in specific angular momentum in the Hydro+ runs. 

\begin{figure*}
	   \center{\includegraphics[width= .9 \textwidth]
	       {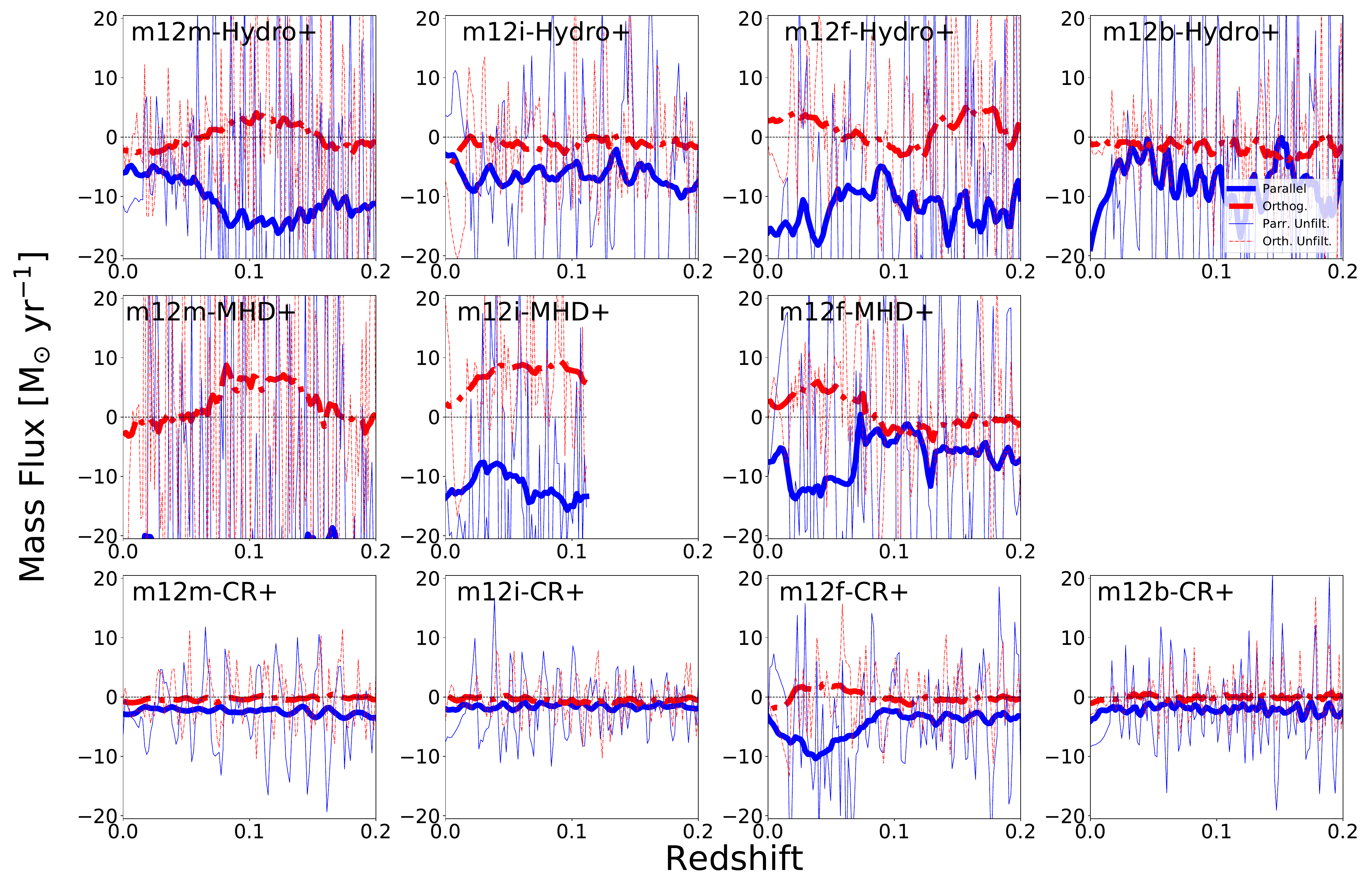}}
	  \vspace{.1 in}
	  \caption{\label{fig:comp_mf_directions} Mass fluxes parallel and orthogonal to galactic disks. Negative values represent in-flowing material. Parallel mass fluxes (blue) are measured within 1 scale height of the disk through a 0.5 kpc radial bin centered at 0.5 \rgalStop. Orthogonal mass fluxes (onto the disk plane; red) are measured within 0.5 \rgal in a 0.5 kpc vertical bin at 1 scale height above the disk. Both \rgal and scale heights were calculated at each snapshot to account for disk growth over time. Thick lines are smoothed using a moving average filter with a window width equal to the dynamical time for each galaxy in order to suppress oscillations. Unfiltered values are shown by thin lines. In general, the parallel flows in all runs are a more consistent source of inflow than the orthogonal flows. The MHD+ and Hydro+ runs tend to be more chaotic largely due to higher star formation rates.
	  }
\end{figure*}

Fig.~\ref{fig:comp_mf_directions} shows the relative contributions of flows parallel and orthogonal to the disk. As in the CR+ runs, the parallel flows are more consistently inflowing in the Hydro+ and MHD+ runs. They are more chaotic though, largely due to the higher star formation rates.

 If observable, the more extreme radial velocities present in the Hydro+ and MHD+ runs would be easier to detect, however, they do not compare as well to observed galactic properties. The star formation rates within these galaxies are slightly higher than expected \citep{hopkins18}, fluctuate more significantly over time, and have relatively higher velocity dispersion in the ISM gas (Chan et al., in preparation). We therefore suggest that detailed findings in the CR runs are better matched to observed $L_*$ galaxies.

\begin{table*}
\centering
\begin{tabular}{p{1.75cm} p{1.75cm} p{1.75cm} p{1.75cm} p{1.75cm} p{1.75cm} p{1.75cm}   }
\hline
Simulation & Run & \rgal & $M_{*}$ & Total Accretion & Disk Mass Flux & SFR \\
Name & & [kpc] & [$M_{\odot}]$ & [\MFunit] & [\MFunit] & [\MFunit] \\
\hline
\hline
\bf{m12m}  & Hydro+ & 17.1 & 1e11 & 5.6 $\pm$ 6.7 & 11 $\pm$ 2 & 15 $\pm$ 5 \\
& MHD+ & 11.7 & 1e11 & 6.6 $\pm$ 9.9 & 18.5 $\pm$ 37.0 & 8.9 $\pm$ 2.0 \\
& CR+ & 26.8 & 3e10 & 1.7 $\pm$ 0.5 & 2.7 $\pm$ 0.4 & 2.5 $\pm$ 0.6 \\
 
\hline

\bf{m12i} & Hydro+ & 24.7 & 7e10 & 6.8 $\pm$ 3.2 & 6.1 $\pm$ 0.9 & 6.6 $\pm$ 3.5 \\
& MHD+ & 20.6 & 7e10 & 3.0 $\pm$ 3.4 & 13.7 $\pm$ 17.5 & 6.5 $\pm$ 2.0\\
& CR+ & 17.1 & 3e10 & 2.2 $\pm$ 1.0 & 1.5 $\pm$ 0.3 & 1.7 $\pm$ 1.5 \\
 
\hline

\bf{m12f} & Hydro+ & 29.3 & 8e10 & 12.5 $\pm$ 15.9 & 9.2 $\pm$ 1.2 & 7.5 $\pm$ 3.4 \\
& MHD+ & 30.1 & 8e10 & 9.8 $\pm$ 6.4 & 11.2 $\pm$ 23.9 & 7.1 $\pm$ 1.6 \\
& CR+ & 18.3 & 4e10 & 3.7 $\pm$ 10.2 & 4.4 $\pm$ 1.6 & 2.3 $\pm$ 2.0 \\
 
\hline

\bf{m12b} & Hydro+ & 18.7 & 1e11 & 6.2 $\pm$ 5.7 & 6.7 $\pm$ 1.3 & 5.4 $\pm$ 1.6 \\
& CR+ & 11.7 & 4e10 & 2.9 $\pm$ 1.9 & 1.8 $\pm$ 0.7 & 1.9 $\pm$ 1.3 \\

\hline

\end{tabular}

\caption{\label{table:mf-comp} Comparison between \rgalStop, stellar mass, total accretion, mass flux through the disk, and total SFR for Hydro+ runs, runs with MHD but no additional CR physics (MHD+), and CR+ runs (presented in the main paper). Accretion, mass flux, and SFR values were averaged from redshift z = 0.2 to 0. Values were calculated as described in Table~\ref{table:disks} and Table~\ref{table:ba-mf}. Accretion, mass flux, and SFR values are $\sim$2 and 4 times higher in the Hydro+ and MHD+ runs, consistent with higher stellar masses.}
\end{table*}

\newpage

\section{Misc. Figures}\label{sec:appendix_II}
\begin{figure}
    \centering
    \includegraphics[width= .5 \textwidth]{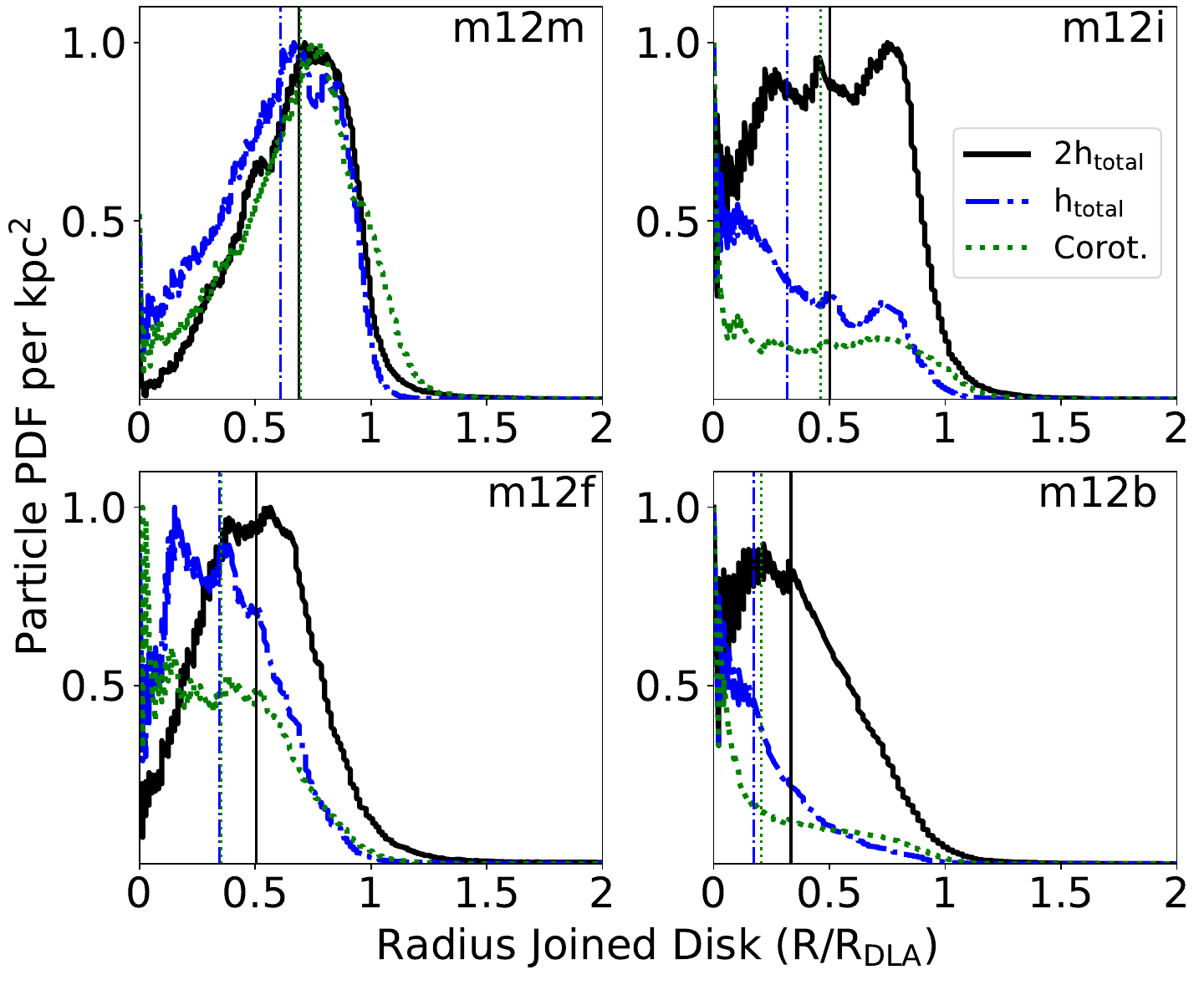}
    \caption{Version of Fig.~\ref{fig:pt-join} but showing accretion per unit face-on area as a function of galactocentric distance in the disk plane. We again show results for three different accretion definitions based on the scale-height of the gas and level of rotational support. Vertical lines show the median of the distribution. While radial distribution of gas accretion is clearly biased towards infall in the outskirts of galaxies (see Fig.~\ref{fig:pt-join}), accretion per unit area  is more flat.
   Compact disk of {\bf m12b} exhibit more centrally concentrated distribution, while the most extended disk {\bf m12m} still shows accretion peaked in the outskirts.}
    \label{fig:pt-rJoin-areaWeight}
\end{figure}

    This appendix shows various supplementary figures to the main text.  Fig.~\ref{fig:pt-rJoin-areaWeight} is a histogram of where the accreting particle sample (see Section~\ref{sec:analysis-pt}) joins the disk weighted by area. It is the same as Fig.~\ref{fig:pt-join}, but each bin is divided by the area of its annulus. This naturally flattens the curve in all cases, in broad agreement with semi-analytic models of galactic disk evolution that require such flat distributions to match a broad range of disk scaling relations \citep{forbes19}, although \textbf{m12m} still peaks near the disk edge. 
    
\begin{figure}
\center

	   \center{\includegraphics[width=.48\textwidth]
	       {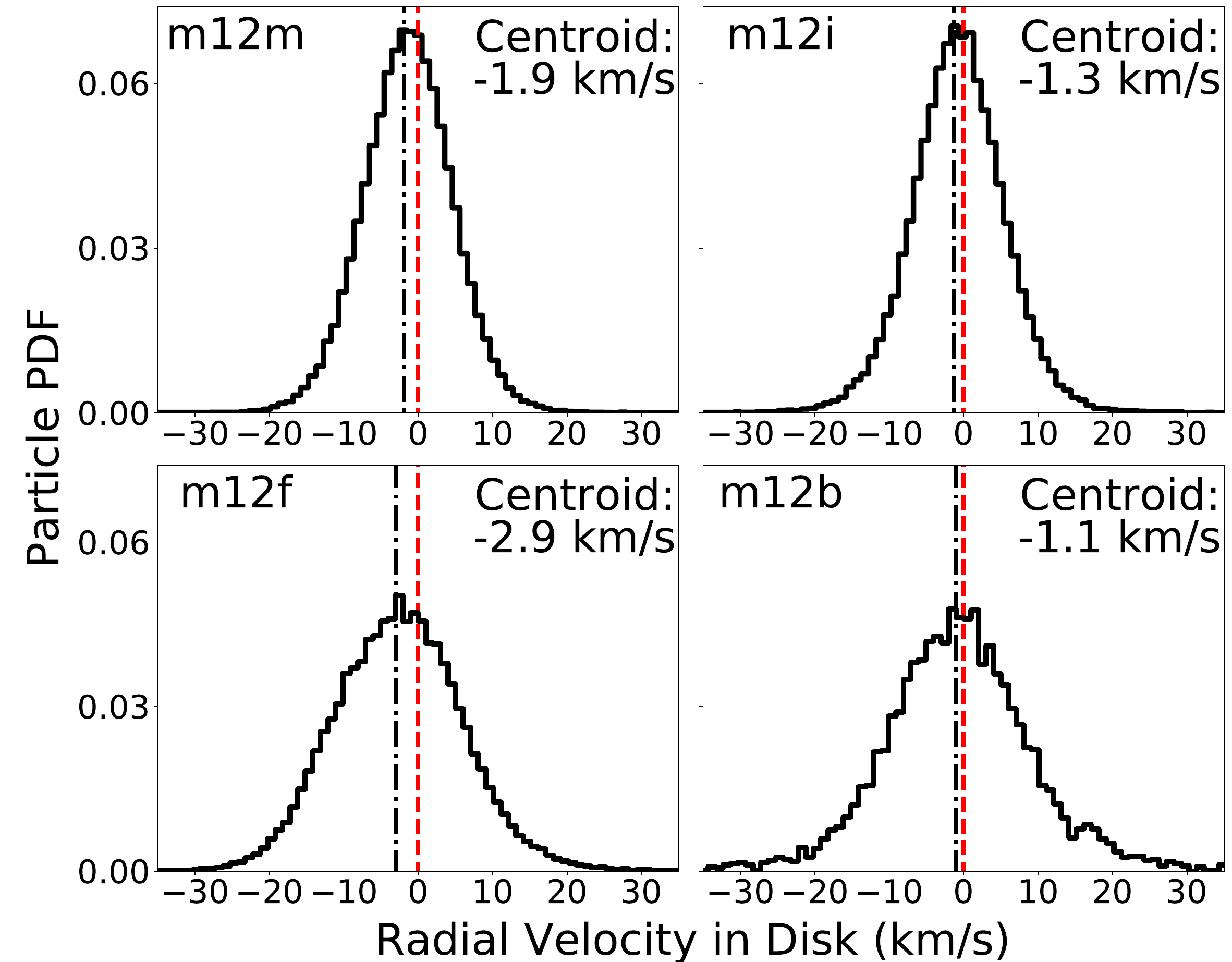}}
	  \caption{\label{fig:pt-diskMovement}  Radial velocities of a selection of gas particles within 0.25\rgalStop$\leq$r$\leq$\rgalStop. Velocities for individual gas elements were averaged over 1 dynamical time (Table~\ref{table:disks}) in order to reduce the effects of spiral arm motion. Velocities of all gas elements are biased inwards and are on average 1-3 km/s.
}
\end{figure}

    Fig.~\ref{fig:pt-diskMovement} is a histogram that shows the radial velocities of individaul gas elements within the disk averaged over time. This gives the same average values as the azimuthal averaging (Fig.~\ref{fig:ba-radvel_and_radmf}), although some particles move with average velocities up to 20 km/s.
    
\begin{figure}
    \centering
    \includegraphics[width=.48\textwidth]{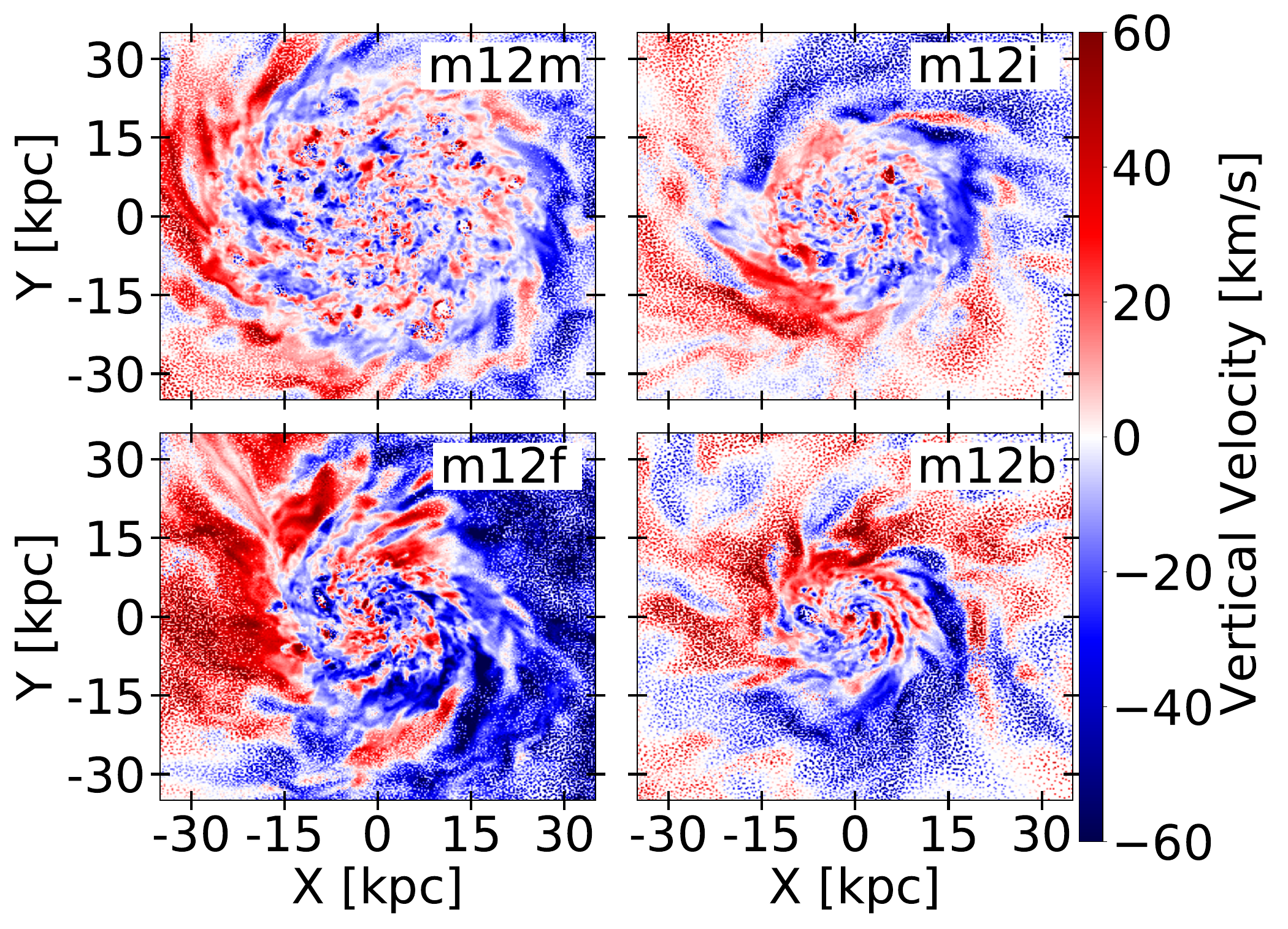}
    \caption{\label{fig:ba-verticalVelMaps} Face on views showing averaged vertical velocities for all gas within 1.5 kpc of the disk plane at z=0. For corresponding parallel velocities, see Fig.~\ref{fig:radvel-maps}.}
    \label{fig:vertical_velocity_maps}
\end{figure}

    Fig.~\ref{fig:vertical_velocity_maps} shows the face on average vertical velocities (orthogonal to the disk). Blue/negative values represent inflow both above and below the disk, while red/positive values represent outflows (i.e. velocities of particles below the disk were multiplied by -1 to consistently represent inflow/outflow when averaged together). Of particular note, we can see small regions of large outflows near the disk centers, particularly in \textbf{m12i} and \textbf {m12f}, as would be expected from concentrated stellar feedback.

\newpage

\section{Resolution Comparisons}\label{sec:appendix_res_comp}

\begin{figure}
	   \center{\includegraphics[width= .5 \textwidth]
	       {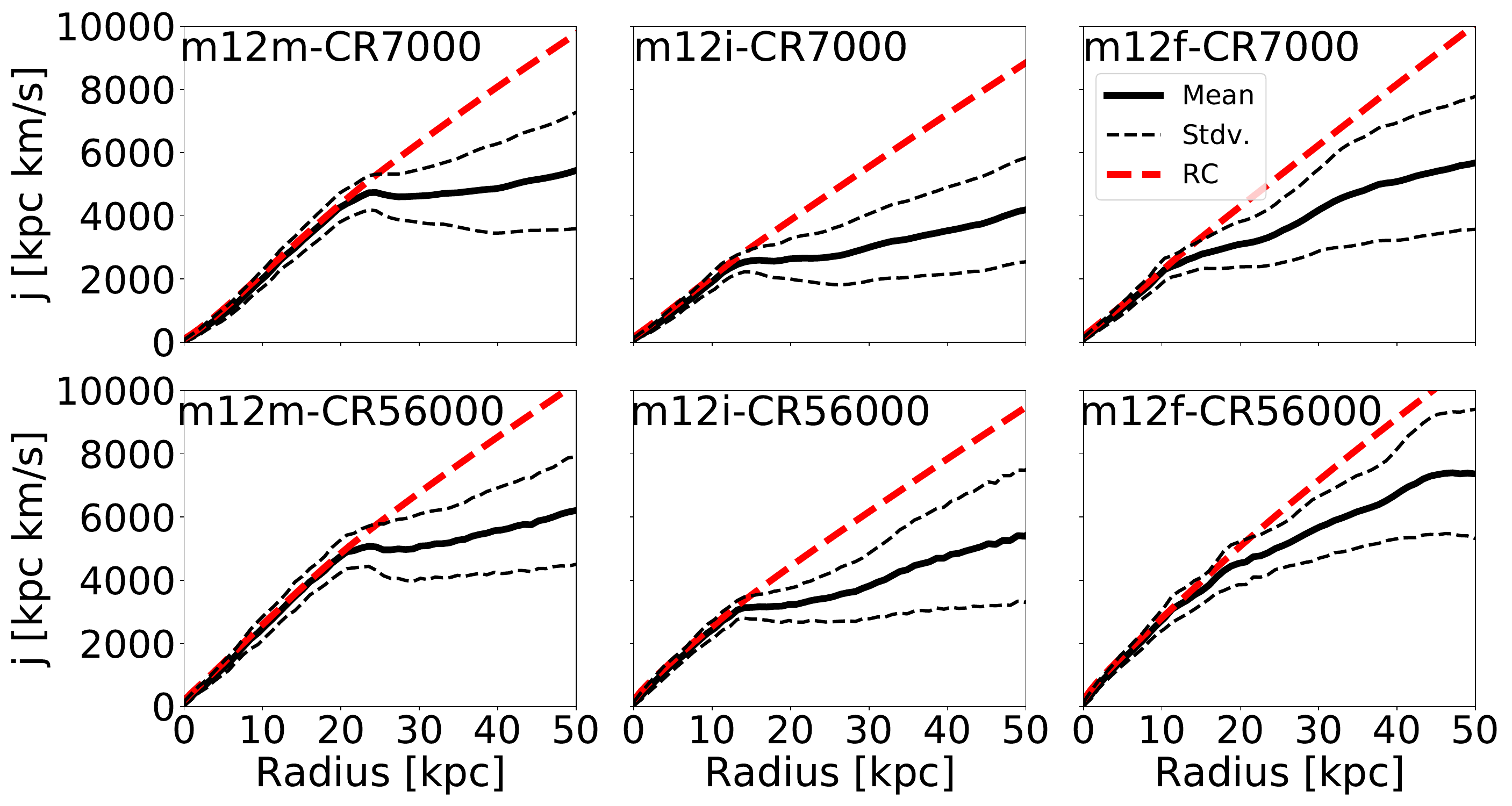}
	  \vspace{.1 in}
	  \caption{\label{fig:angMom_resComp} Comparison between mass-weighted average specific angular momentum curves as a function of spherical radius between res7000 and res56000 runs. Results are qualitatively similar.}
}
\end{figure}

This section briefly compares a lower resolution CR+ simulation run with original gas particles mass of $m_{\rm gas} \sim 56000 \msun$ (cr56000) to the runs used as default in this paper with $m_{\rm gas} \sim 7000 \msun$ (cr7000). Fig.~\ref{fig:angMom_resComp} compares how the specific angular momentum of gas changes with radius. Curves are very similar in all cases, following the rotation curve tightly within the disk, and flattening outside the disk. Overall we do not see any strong resolution trends in qualitative and quantitative nature of gas infall close to galactic disks.

\end{document}